\documentclass{aa}
\usepackage{txfonts}
\usepackage{graphicx}
\usepackage{natbib}
\bibpunct{(}{)}{;}{a}{}{,} 

\begin{document}


\title{Spitzer spectral line mapping of the HH211 outflow}

\author{O. Dionatos\inst{1,2}
          \and
          B. Nisini\inst{2}
          \and
          S. Cabrit\inst{3}
          \and
          L. Kristensen\inst{4}
          \and
          G. Pineau des For{\^e}ts\inst{5,3}
          }

\institute{
		Natural History Museum, University of Copenhagen, {\O}ster Voldgade 5-7 1350 Copenhagen, Denmark \\
              \email{odysseas@snm.ku.dk}\\
         \and
                INAF - Osservatorio Astronomico di Roma, Via di Frascati, 33  00040 Monte Porzio Catone (RM), Italy\\
              \email{nisini@oa-roma.inaf.it}\\
         \and
                LERMA, Observatoire de Paris, UMR 8112 du CNRS, 61 Avenue de Observatoire, 75014 Paris, France\\
              \email{sylvie.cabrit@obspm.fr}\\
         \and
                Leiden Observatory, Leiden University, Niels Bohrweg 2, 2300 CA Leiden, The Netherlands\\
			\email{kristensen@strw.leidenuniv.nl}\\
	\and
		Institut Astrophysique Spatiale (IAS), UMR 8617, CNRS, Universite Paris-Sud 11, Batiment 121, 91405 Orsay Cedex, France\\
			\email{guillaume.pineaudesforets@ias.u-psud.fr}
             }

\abstract
{Jets from the youngest protostars are often detected only at mm wavelengths, through line emission of CO and SiO. However, it is not yet clear if such jets are mostly molecular or atomic, nor if they trace ejected gas or an entrained layer around an embedded atomic jet.}
%
{We investigate the warm gas content in the HH211 protostellar outflow to assess the jet mass-flux in the form of H$_2$ and probe for the existence of an embedded atomic jet.}
{We employ archival Spitzer slit-scan observations of the HH211 outflow over 5.2 -- 37 $\mu$m obtained with the low resolution IRS  modules. Detected molecular and atomic lines are interpreted by means of emission line diagnostics and an existing grid of molecular shock models.  The physical properties of the warm gas are compared against other molecular jet tracers and to the results of a similar study towards the L1448-C outflow.}
{We have detected and mapped the v=0--0 S(0) - S(7) H$_2$ lines as well as fine-structure lines of
S,  Fe$^+$, and Si$^+$. The H$_2$ is detected down to 5$\arcsec$ from the source and is characterized by a "cool" T$\sim$ 300K and a "warm" T$\sim 1000 \pm 300$~K component, with an extinction$A_V \sim 8$ mag. The amount of cool H$_2$ towards the jet agrees with that estimated from CO assuming fully molecular gas. The 
warm component is well fitted by C--type shocks with a low beam filling factor $\sim$ 0.01-0.04 and a mass-flux similar to the cool H$_2$. The fine-structure line emission arises from dense gas with ionization fraction $\sim 0.5 - 5\times 10^{-3}$, suggestive of dissociative shocks. Line ratios to sulfur indicate that iron and silicon are depleted compared to solar abundances by a factor $\sim$ 10--50.}
{Spitzer spectral mapping observations reveal for the first time a cool H$_2$ component towards the CO jet of HH211 consistent with the CO material being fully molecular and warm at $\simeq$ 300 K. The maps also reveal for the first time the existence of an embedded atomic jet in the HH211 outflow that can be traced down to the central source position. Its significant iron and silicon depletion excludes an origin from within the dust sublimation zone around the protostar. The momentum-flux seems insufficient to entrain the CO jet, although current uncertainties on jet speed and shock conditions are too large for a definite conclusion. 
}


\maketitle


\section{Introduction\label{sec1}}

The process of mass accretion leading to the formation of low mass protostars is always associated with ejection of material in the form of well collimated high-velocity jets and/or wider slower bipolar outflows \citep{Reipurth,Arce}. Accretion and ejection phenomena are believed to be intimately connected through the existence of a magnetized accretion disk around the forming protostar, but there is no consensus so far about the exact ejection mechanism  \citep{Pudritz, Shang,Ferreira}.  The study of the physical properties of jets (e.g. temperature, density, abundances, mass flux) is therefore crucial to obtain indirect constraints on the mass accretion  and ejection processes, especially in the case of protostars in the earliest evolutionary stages - the so called Class 0 protostars - where the launching zone is heavily embedded in a dense cocoon of dust and gas.   

Due to this high extinction,  jets from the youngest Class 0 sources cannot be seen in the optical and are mostly traced in near-IR H$_2$ and in mm CO and SiO lines \citep{Guilloteau,MacC,Davis}. However, it is still unclear whether the molecules trace a sheath of entrained ambient material around an unseen underlying atomic jet, or if they trace the primary jet material itself. In the latter case, the chemical composition of the jet (molecular vs. atomic fraction, dust depletion) holds crucial clues to the ejection zone, as the strong UV flux generated by the accretion shock should destroy dust grains and H$_2$ molecules in the innermost regions of the disk. A wind
from such inner zones would be mostly atomic even if CO and SiO are abundant --- unless the mass flux is unusually large \citep{Ruden}. 

Mid-infrared spectroscopic observations are essential to progress on these questions. A Spitzer study of the innermost jet regions of the Class 0 L~1448 outflow \citep{Dionatos} revealed an underlying deeply embedded atomic/ionic component, not seen in the near-IR range, as
well as mid-IR H$_2$ emission tracing the warm molecular content. Here we present a more extensive Spitzer mapping study of the HH211 outflow, a particularly interesting target that appears as a "text-book" example of jet-driven flow.

HH211 is located in the IC~348 complex in Perseus at a distance estimated to be between 250 pc \citep{Enoch},  adopted in the current paper, and
$\sim$320 pc \citep{Herbig, Lada}. The driving source HH211-mm is  detected in mm continuum emission \citep[e.g.][]{Gueth, Lee}, and has been classified as a low-mass and low-luminosity Class 0 young stellar object \citep{Froebrich}. 

The outflow was discovered by \citet{MacC} in near-IR H$_2$ emission tracing 2000~K hot, shocked gas in two 
bright symmetric bowshocks separated by 0.13pc with a faint chain of knots in between. 
A few compact knots of atomic emission at optical and near-IR wavelengths (i.e. 
H$_{\alpha}$, [SII]6730\AA\, and [FeII]1.64$\mu$m) are also seen 
near the brightest H$_2$ peaks \citep{Oconnell,caratti,walaw2,walaw1}.

The absence of any further shock structures on larger scales makes it the youngest outflow known so far, with a kinematical age of only 1000 yr $\times (V/100)$ km~s$^{-1}$ \citep{Eisloeffel,Gueth}. 

Low velocity CO J=2-1 \citep{Gueth} and CO J=3-2 \citep{Lee} observations delineate the shape of
a bipolar swept-up cavity with its ends tracing the H$_2$ bow-shocks and its flanks connecting back to the driving source, while the high-velocity components of the same transitions reveal an inner well-collimated CO jet extending out to $\pm 25\arcsec$ from the source.
The CO jet is also traced in SiO through interferometric observations of the  J=1-0 \citep{Chandler}, 5-4 \citep{Hirano} and  8-7 \citep{Lee} transitions, and single-dish multi-transition studies \citep{Gibb, Nisini1}. Excitation of such high $J$ levels implies that the SiO jet is warmer and denser than the outflow cavity traced by low-velocity CO emission. 

All observations reveal a highly bipolar structure, with blue and red-shifted lobes pointing southeast and northwest from the driving source. The two lobes are well separated and such morphology combined with the low observed radial velocities \citep[$\sim$20 km s$^{-1}$,][]{Gueth} implies that the inclination of the outflow is less than 10$^o$ from the plane of the sky \citep{Gueth, Chandler, Lee}. This simple geometry is favorable to modelling. Indeed,  
the CO cavity shape is well fitted by  a dynamical model of jet-driven bowshock propagating into a medium of decreasing density \citep{Gueth}.  This interpretation is supported by fitting 
the near-IR line fluxes and morphologies at the apex of the redshifted lobe with a series of bowshocks \citep{Oconnell}. A Spitzer map obtained at the tip the blue lobe indicates a higher excitation 40 km/s dissociative shock on that side, with a substantial UV flux \citep{Tappe}. 

In this paper, we present archival Spitzer-IRS slit-scan observations covering most of the HH211 flow with
low spectral resolution (Section 3), which we analyse to address the physical conditions and the molecular and dust content. In Section 4, the detected rotational H$_2$ transitions are employed as probes of the physical conditions in the jet close to the driving source and the shocked gas further out, and compared  with an existing grid of C and J-type shock models. Atomic lines present in our spectra are treated separately as probes for the existence of a deeply embedded atomic jet, and used to measure dust depletion. Mass-fluxes are inferred for the H$_2$ and atomic components, and compared to those measured in CO and SiO, as well as to a similar study of the inner jet of L1448. Our conclusions are presented in Section 5.


\begin{figure}
\centering
\includegraphics[scale=0.44, angle=0]{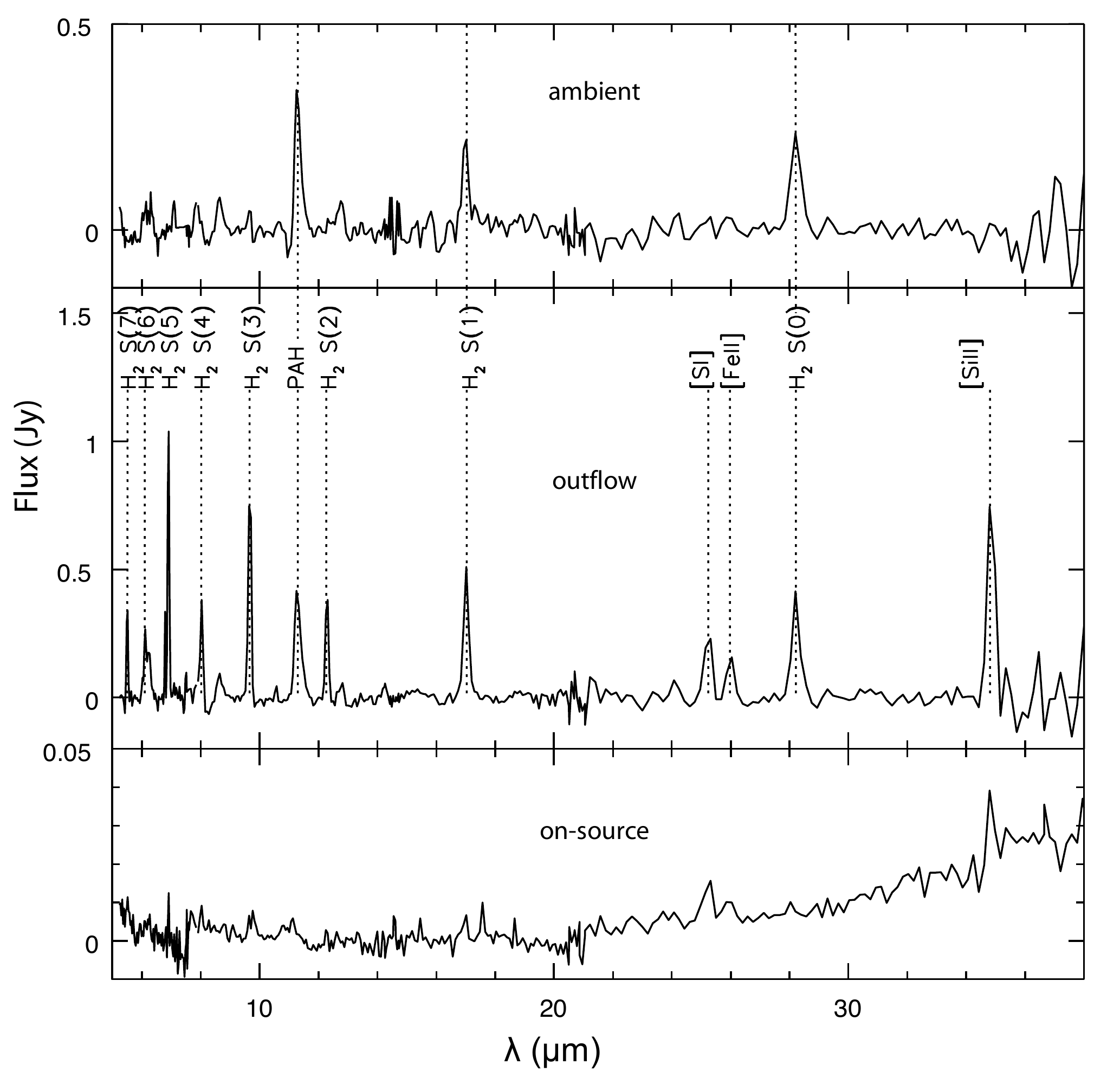}
\caption{Continuum subtracted spectrum, integrated over a region of 612 square arcseconds covering the HH211 flow (\textit{middle}),
shown in comparison with a spectrum of the off-outflow region of the same area (\textit{top}). H$_2$ emission from the 0--0 S(0)-- S(7) lines is detected in the outflow, together with atomic emission from the ground state transitions of
[FeII],[SiII] and [SI]. The H$_2$ S(0) and S(1) lines are detected also in the off-outflow spectrum, 
indicating the presence of a diffuse line emission component not related to the HH211 flow. Diffuse emission
from PAH features, such as the 11.3$\mu$m feature, is also evidenced. (\textit{Bottom:}) On source spectrum extracted from an area equal to the LL pixel scale (110.25 square arcseconds), after removing the zodiacal light and ambient contribution presenting the intrinsic continuum emission from the outflow source.}
\label{fig0}
\end{figure}


\begin{figure*}[!t]
\centering
\includegraphics[width=12cm, angle=0]{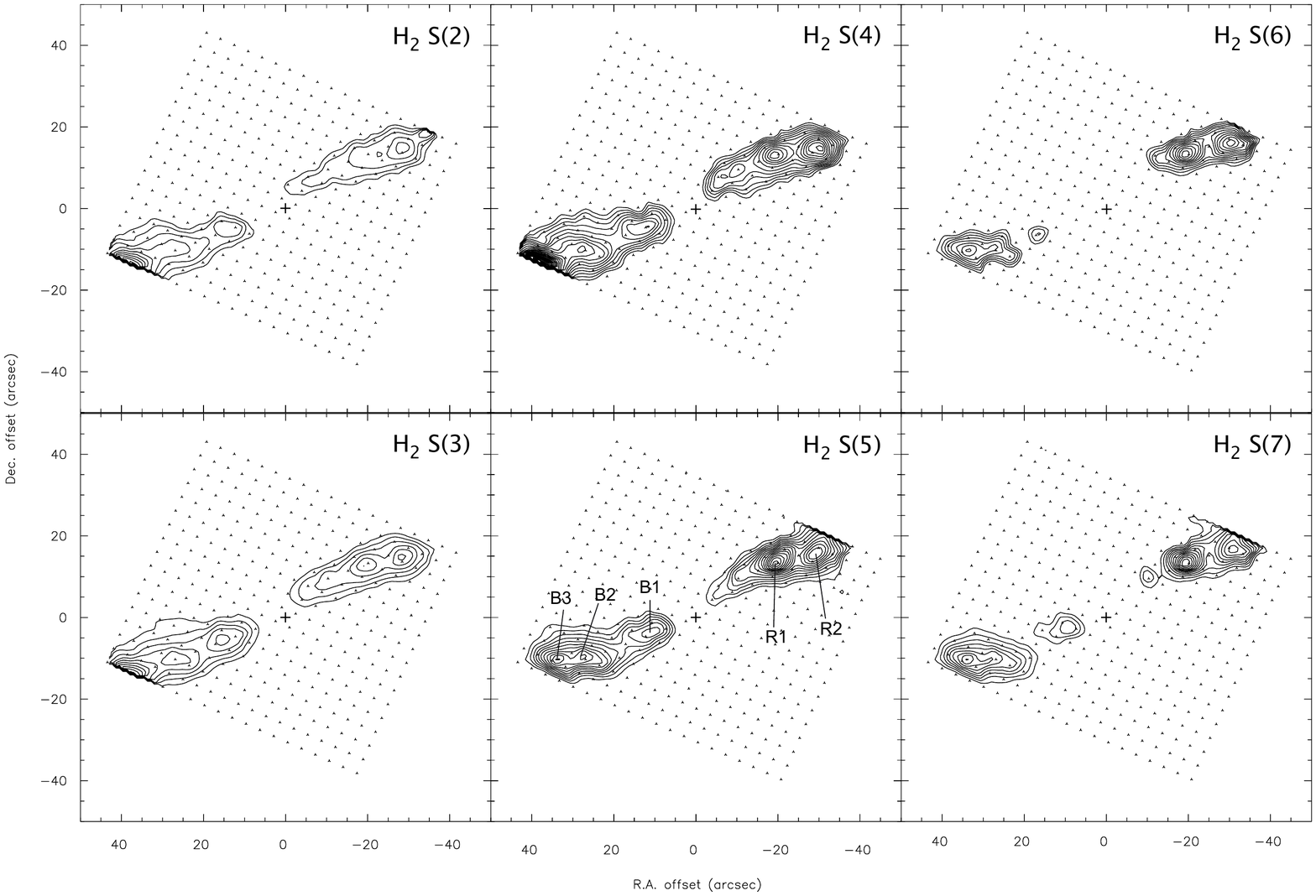}
\caption{H$_2$ v=0--0 S(2) -- S(7) line intensity maps of the HH211 flow observed with the {\it Spitzer} SL module, reconstructed with a pixel scale of 3.5$\arcsec$. Contours are from 10$^{-12}$ W cm$^{-2}$ sr$^{-1}$ with an increment of 8 10$^{-13}$ W cm$^{-2}$ sr$^{-1}$ for the para lines (even $J$), and from 2 10$^{-12}$ W cm$^{-2}$ sr$^{-1}$ with an increment of 2.5 10$^{-12}$ W cm$^{-2}$ sr$^{-1}$ for the ortho transitions (odd $J$). Crosses indicate the driving source position at $\alpha_{J2000}$:03h43m56.5s, $\delta_{J2000}$:+32d:00m:51s. In the H$_2$ S(5) map, points B1-B3 and R1, R2 indicate peaks of emission, where A$_V$ and ortho to para ratio are estimated and excitation analysis is carried out.}
\label{fig1}
\end{figure*}

\section{Observations and data reduction \label{sec2}}

Observations were  obtained with the \textit{Spitzer} satellite \citep{Werner}
and retrieved from the Spitzer Public Data Archive using the Leopard software.  They
were performed as part of the "Shock dissipation in Nearby Star Forming Regions" program conducted by J. Bally (P.I.). In these, the low resolution modules (R$\sim$60-130) of the \textit{Spitzer Infrared Spectrograph} \citep[IRS,][]{Houck} were used in slit-scan mode to cover an area of 57$\arcsec$ $\times$ 73$\arcsec$ and 157$\arcsec$ $\times$ 168$\arcsec$ for the Short Low (SL) and Long Low (LL) modules respectively, centered on the Class 0 source HH211-mm ($\alpha_{J2000}$:03h43m56.5s, $\delta_{J2000}$:+32d00m51s). A perpendicular scan step equal to the slit width of each module was used, with a total integration time of 171 min. The combination of both IRS low resolution modules gives a complete wavelength coverage of 5.2 -- 37.0 $\mu$m. 

Initial data processing was performed at the Spitzer Science Center using Version 15 of the processing pipeline. Spectral data-cubes were built using the CUBISM software package \citep{Smith} and bad/rogue pixels were masked through visual inspection. 

Figure \ref{fig0} presents the extracted continuum-subtracted spectrum, integrated over a region encompassing
 the HH211 flow. The full series of H$_2$ pure rotational lines (S(0)-S(7)) were detected,  along with atomic  and ionic lines from the fundamental transitions of [SI], [FeII], and [SiII] at 25$\mu$m, 26$\mu$m, and 35$\mu$m  respectively. For comparison, Fig. \ref{fig0} shows also a spectrum of a region encompassing the same
 area but in a direction perpendicular to the outflow. The H$_2$ S(0) and S(1) lines are detected also in this off-outflow spectrum, 
indicating the presence of a diffuse line emission component not related to the HH211 flow. Diffuse emission
from PAH features, such as the 11.3$\mu$m feature, is also evidenced. In Figure \ref{fig0} we also plot the on source spectrum for an area equal to the LL pixel size, after subtracting the contribution from an off-source, free of line-emission position. This only removes zodiacal light and reveals the
intrinsic mid-IR continuum from the central source, which is typical of low-luminosity Class 0 sources.

Subsequent analysis consisted in the construction of individual line emission maps, using a home-built pipeline;
in this, for each spatial pixel of the data-cube, the brightness of each spectral line of interest was calculated by Gaussian fitting after subtracting a  local second order polynomial baseline.  The resulting line intensity maps have a square pixel of side equal to the slit width of the IRS module, namely 3.5$\arcsec$ and 10.5$\arcsec$ for the SL and LL modules respectively, while the diffraction limit of the telescope is 2.4$\arcsec$ and 6.0$\arcsec$ at $\lambda$=10$\mu$m and 25$\mu$m respectively. The astrometric accuracy of the maps was found to be good within the limits imposed by the pixel size of each IRS module.  
\section{Spectral line maps \label{sec3}}

\subsection{H$_2$ emission}

Figure \ref{fig1} presents the emission line maps of the S(2) - S(7) rotational H$_2$ lines observed with the SL - IRS modules with a 3.5$\arcsec$  sampling. Contours shape a characteristic bipolar outflow pattern, where the H$_2$  emission is detected down to a projected angular distance $\sim$ 5\arcsec\ from the driving source. Further downwind, peaks of emission which can be attributed to shocked gas are observed. In the H$_2$ S(5) map of Fig. \ref{fig1} we label these peaks as B1-B3 in the southeast, blue-shifted lobe and R1-R2 in the northwest, red-shifted lobe, respectively.

\begin{figure}
\centering
\includegraphics [scale=0.48, angle=0] {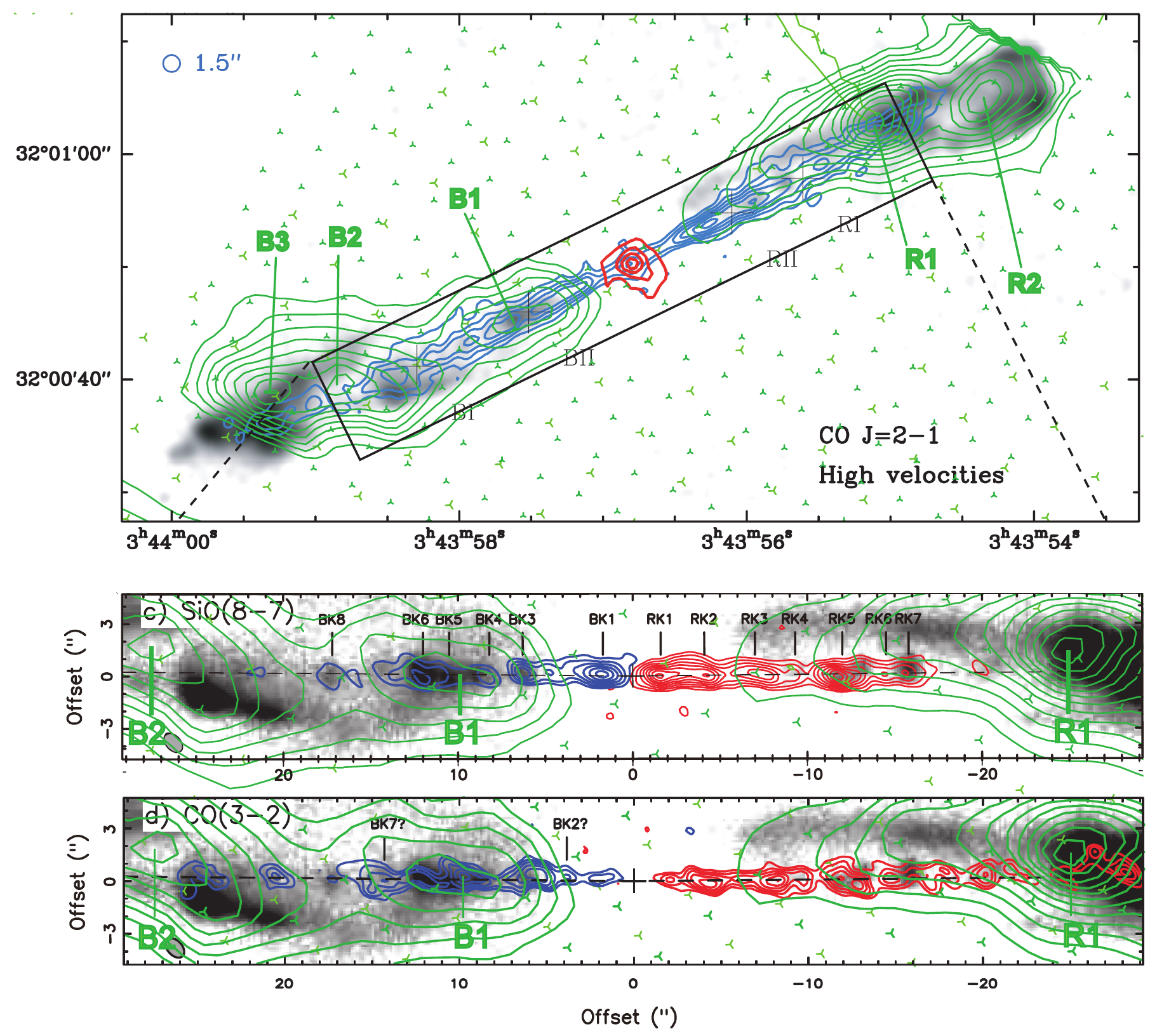}
\caption{\textit{(Ipper panel)} H$_2$ v=0--0 S(5) map (green contours) superimposed over the high velocity CO \textit{J} = 2--1 map of \citet{Gueth}; \textit{(Lower panels)} Inner part of the H$_2$ v=0--0 S(5) map (green contours) superimposed  on the SiO \textit{J} = 8--7 and the CO \textit{J} = 3--2 maps of \citet{Lee}.  The grayscale backgrounds show the H$_2$ v=1--0 S(1) image from \citet{MacC}(top panel) or \citet{Hirano}(lower panels). The slight S-shape pattern of the S(5) emission map follows the brightness asymmetry of the cavity seen in the 2.12 $\mu$m image. In the blue-shifted jet lobe, point B1 is coincident with peak BII in CO $J$=2-1 and BK4--6 in SiO.}
\label{fig2}
\end{figure}

Figure \ref{fig2} shows the H$_2$ S(5) emission map overlaid on the high-velocity CO \textit{J} = 2$\rightarrow$1 map of \citet{Gueth} (upper panel) and on the SiO \textit{J} = 8$\rightarrow$7 and CO \textit{J} = 3$\rightarrow$2 maps of  \citet{Lee} (lower panels). The grayscale background shows the near-IR  H$_2$ v=1-0 S(1) images from \citet{MacC} or \citet{Hirano}. Overall, the spatial correspondence between mid-IR and near-IR H$_2$ lines is quite good except towards the broadest parts of the outflow cavity, where midIR emission appears to trail behind the near-IR one.

In the southeastern - blue lobe, the peak of mid-IR H$_2$ emission closest to the source (B1) coincides with the 2.12$\mu$m H$_2$ knot G in the nomenclature of \citet{MacC}, the BII knot of the CO jet \citep{Gueth} and knots BK4-6 in the SiO jet \citep{Lee} (see lower panel of Fig.  \ref{fig2}). Therefore it appears to trace mostly the jet beam. 
Further outwards, Peak B2 is centered in a hole {\it behind} the bowshock rim delineated in the near-IR image, while Peak B3 appears to lie on the bow rim (note however that the spatial extent of our SL map does not fully cover the tip of this region).

Symmetrically on the redshifted side, the mid-IR H$_2$ emission close to the source forms an extended curving "finger"  tracing both the jet and the cavity wall north of it, as seen in the near-IR (see Fig. \ref{fig2}) and peak R1 coincides with the near-IR knot F of \citet{MacC}. However, peak R2 is again centered on the cavity behind the bright bow rim traced in the near-IR. The spatial offset between B2 and R2 with respect to the near-IR counterparts suggests that mid-IR emission in these broad regions may be tracing the outer bow-shock wings where gas is expected to be in lower excitation conditions. 

\begin{figure*}
\centering
\includegraphics[width=14cm, angle=0]{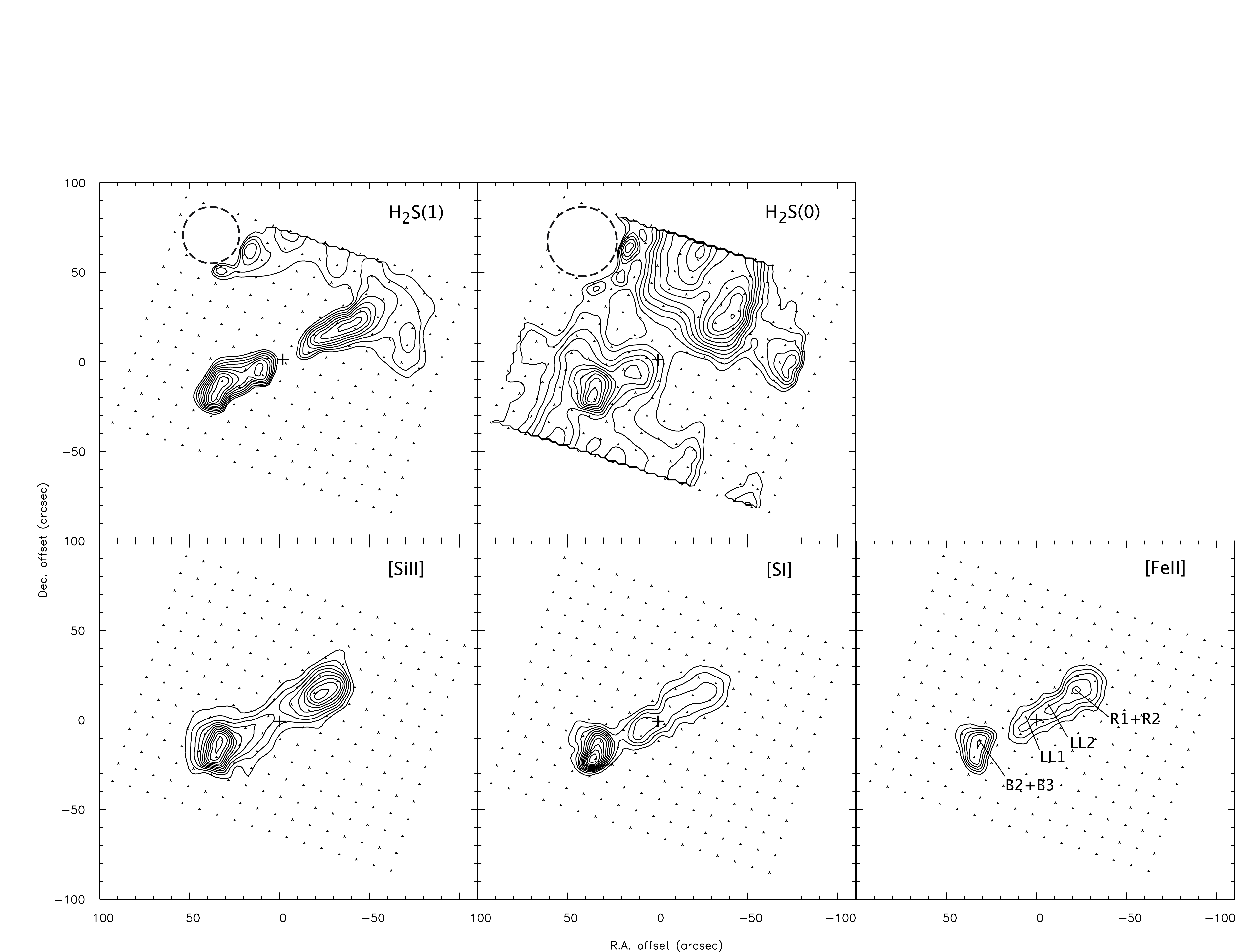}
\caption{Intensity maps of the HH211 flow in the H$_2$S(1) and S(0) lines ({\it top row}) and the atomic [SiII]35$\mu$m, [SI] 25$\mu$m, and [FeII]26$\mu$m lines ({\it bottom row}) 
obtained with the {\it Spitzer} IRS - LL module (pixel scale $\sim$10.5$\arcsec$). Crosses mark the position of the driving source HH211-mm. The dashed circle in the H$_2$ maps masks the area around a bright star (IC348 IR) where line intensity measurements are unreliable. In the [FeII] 26$\mu$m map, contours are interrupted just before the eastern bowshock peak due to a rogue pixel. Contour levels in the H$_2$ S(1) map start from 10$^{-12}$ W cm$^{-2}$ sr$^{-1}$ with 10$^{-13}$ increments, and from 7 10 $^{-13}$ W cm$^{-2}$ sr$^{-1}$ with $5 \times 10^{-14}$ increments for the S(0) map. Contours of the ionic line maps start at 10$^{-13}$ W cm$^{-2}$ sr$^{-1}$ with 10$^{-13}$ W cm$^{-2}$ sr$^{-1}$ steps for [SiII] and [FeII], and at 10$^{-13}$ W cm$^{-2}$ sr$^{-1}$ with $2 \times 10^{-13}$ W cm$^{-2}$ sr$^{-1}$ steps for [SI]. }
\label{fig3}
\end{figure*}

The top row of Figure \ref{fig3} presents the emission maps of the S(0) - S(1) rotational H$_2$ lines observed with the LL - IRS modules with a 10.5$\arcsec$  sampling. Despite the lower angular resolution compared to the SL module, the H$_2$\textit{S}(1) line clearly delineates the same outflow pattern as  in Figure \ref{fig2}. 
As discussed in Sect. 2, faint extended diffuse emission is also present, and becomes mostly evident at the north end of the map; diffuse emission is even more apparent in the H$_2$ S(0) line map of Fig. \ref{fig3}.

This diffuse brightness could be caused by the presence of an extended 
photo-dissociation region (PDR) created at the cloud surface by the illumination of FUV radiation 
from nearby stars. The existence of a diffuse PDR is strongly supported 
 by the detection of diffuse PAH emission, which signals the presence of UV 
radiation in the field. The H$_2$ diffuse emission has a brightness of the order of 
10$^{-5}$ erg/s/cm$^2$ for both the S(0) and S(1) lines: according to the 
\citet{kaufman} model, 
such an intensity
is compatible with a PDR with a density of 10$^4$ cm$^{-3}$ or larger and a FUV field of the
order of 10$^2$ times the average interstellar field, measured in units of G$_0$ (i.e. 
in units of 1.6$\times$10$^{-3}$ erg\,cm$^{-2}$s$ ^{-1}$). 
The most likely candidate for producing the diffuse PDR is the $\o$ Per B0.5 star, 
located at a projected distance of about 15$\arcmin$ north of the HH211 system. 
The FUV field of this star (assuming an effective temperature of 25\,000 K, \citet{hernandez}),
diluted for the distance, can account for a G$_0$ $\sim$ 500, which is compatible with the estimation
given above, taking into considerations also projecting factors and partial field absorption
along the path. 
We cannot exclude that part of the extended S(0) and S(1) H$_2$  emission  also originates from other outflows located
north and south of HH211-mm, seen in both optical and near-IR narrow band imaging of the region \citep{walaw1,eis}. 
However, no close spatial correspondence is evident at our spatial resolution.



\subsection{Atomic emission}

The bottom row of Figure \ref{fig3} presents the emission line maps of
[SiII]35$\mu$m, [SI] 25$\mu$m, and [FeII]26$\mu$m obtained from the LL - IRS modules with a 10.5$\arcsec$  sampling.
The emission from atomic and ionic lines is substantially weaker in comparison to the H$_2$ lines. 
However, unlike H$_2$ lines which are not detected close to the driving source, all three fine structure transition contours are continuous, evidencing emission very close to the protostar HH211-mm. Morphologically, the [SI] and [FeII] maps peak at the blue lobe bow-shock\footnote{In the [FeII] 26$\mu$m map, contours are interrupted just before the actual blue-shifted bowshock peak; this is an artifact resulting from a rogue pixel in the LL-IRS module that had to be masked out since it was introducing unreliable brightness values forming a bright vertical stripe across the map.}. Emission from [SiII] instead shows strong peaks on both the blue and red bow-shocks. 

The enhanced atomic - ionic emission on the blue-lobe bowshock evidences a highly energetic
interaction of the underlying jet with the ambient material. In the same region, a wealth of other molecular and ionic lines such as OH, H$_2$O, HD, [NeII] were mapped with Spitzer by \citet{Tappe}, suggesting a strong UV flux typical of a 40 km $s^{-1}$ dissociative shock. The presence of such high velocity shocks  in this region is also 
consistent with the detection of optical H$_{\alpha}$ and [SII] emission spots \citep{walaw1,walaw2}. In the following, we will show that rotational H$_2$ lines over the HH~211 outflow appear to trace slower shocks, either internal to the jet or in bowshock wings.

\section{Excitation conditions and mass fluxes \label{sec4}}

\subsection{H$_2$ emission \label{sec41}}

\subsubsection{Ortho to para ratio, extinction, and temperature components}

The H$_2$ pure rotational transitions are easily thermalized in protostellar outflow environments owing to their low critical densities ($\sim$10$^2$-10$^4$ cm$^{-3}$) and can be used to probe temperatures in the range between $\sim$300 and 1500 K. The most direct way to estimate the temperature of the emitting gas is  by means of an excitation diagram (EXD); this involves plotting the quantities \textit{ln(N$_{u,J}$/g$_u$)} against \textit{E$_{u,J}$}, where \textit{N$_{u,J}$} and \textit{E$_{u,J}$} are the column density and the energy of the upper level, and $g_u$ the statistical weight including the spin degeneracy \textit{(2S+1)}. Considering LTE conditions, the values of these quantities for the observed H$_2$ transitions should fall on a straight line, from the inverse slope of which the excitation temperature can be derived. Deviations from this linear correlation can be used to constrain other physical properties, such as the H$_2$ ortho-to-para ratio (hereafter, OTP) and the optical extinction A$_V$, as described below.

Deviations of the OTP from its LTE value are reflected as vertical displacements between the ortho and para transitions in the EXD, forming a "saw-tooth" pattern between the two H$_2$ species. The observed OTP was estimated by examining the alignment of the S(5) data point with the neighboring S(4) and S(6) transitions, following the method outlined in \citet{Wilgenbus}; in all cases the spatial OTP variations are small, and values lie very close, within the statistical error limits, to the high-temperature LTE value of 3. 

The H$_2$ S(3) transition at 9.7 $\mu$m is sensitive to the amount of dust along the line of sight, 
as it is located within a wide-band silicate absorption feature at the same wavelength; consequently a visual extinction value can be  estimated by examining the alignment of the S(3) point in comparison to the S(2) and S(4), having previously corrected for any deviations of the OTP ratio from the equilibrium value. Visual extinction is then estimated assuming an  \textit{A$_{9.7}/A_V$} ratio equal to 0.087 \citep{Lebofsky}. $A_V$ is found to exhibit no regular pattern along the outflow and to range between 7 and 9 mag, being consistent with previous estimates of 7--15 mag
by \citet{MacC,caratti}\footnote{\citet{Oconnell} reported a higher extinction of 25 mag in the redshifted lobe but  considered it rather tentative given their observing conditions} .  

\begin{figure*}[!t]
\centering
\includegraphics [width=16cm, angle=0]{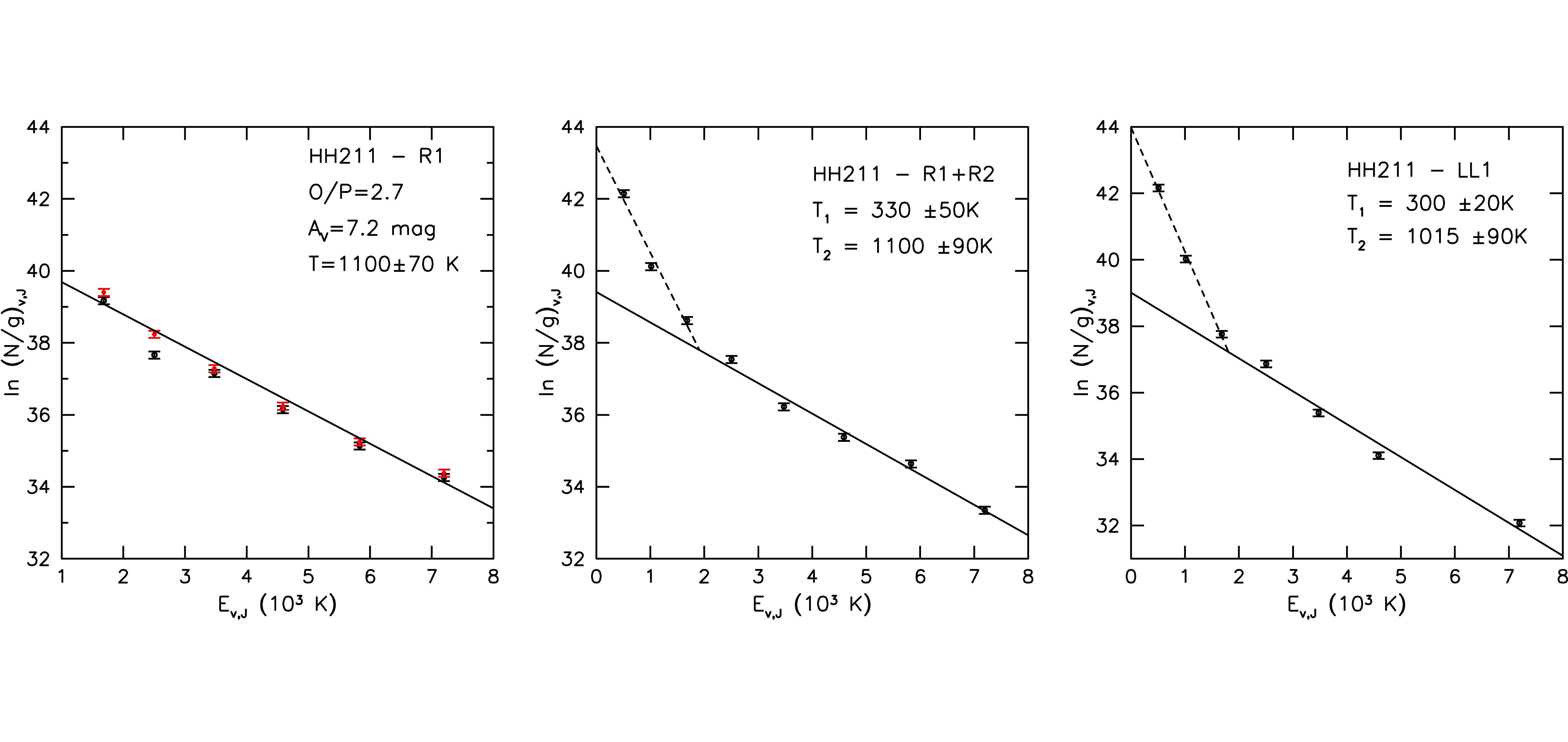}
\caption{Excitation diagrams of H$_2$ in various pixel sizes. (\textit{Left}) H$_2$ columns in peak R1 from transitions in the SL module in 3.5$\arcsec$ pixels (open circles). Ortho to para ratio and extinction are estimated from the alignment of the S(5) and S(3) data points with their neighboring S(2), S(4) and S(4), S(6) respectively. Dereddened columns (filled red circles) are linearly fitted (solid line) to derive the excitation temperature. Derived values are listed on the upper right corner of the diagram. (\textit{Center, right}) Excitation diagrams of peaks R1+R2 and LL1 in 10.5$\arcsec$ pixels, combining the H$_2$ transitions observed in both the LL and SL modules. The SL line maps were re-extracted at the LL map pixel scale. A second cooler component  is apparent in S(0) -- S(2) with $T_1\sim 300$~K; the higher H$_2$ transitions trace a warmer component with temperature $T_2 \sim 1100$~K identical to that inferred at the SL pixel scale (see left panel).} 
\label{fig4}
\end{figure*}

Considering an average value of A$_V$ equal to 8 mag  across the map, we have dereddened all the detected H$_2$ lines falling in the SL range (ie. S(2) to S(7)) using the extinction law of \citet{Lebofsky}. Consequently, for all the points of the SL data cube where at least 3  H$_2$ lines are detected above 3-sigma, we have calculated the excitation temperature from the slope of a least square fit on the dereddened data points in the EXD.  An example of these operations is shown in the left panel of Fig. \ref{fig4} in the case of the R1 peak. 

\begin{figure}
\includegraphics[scale=0.6]{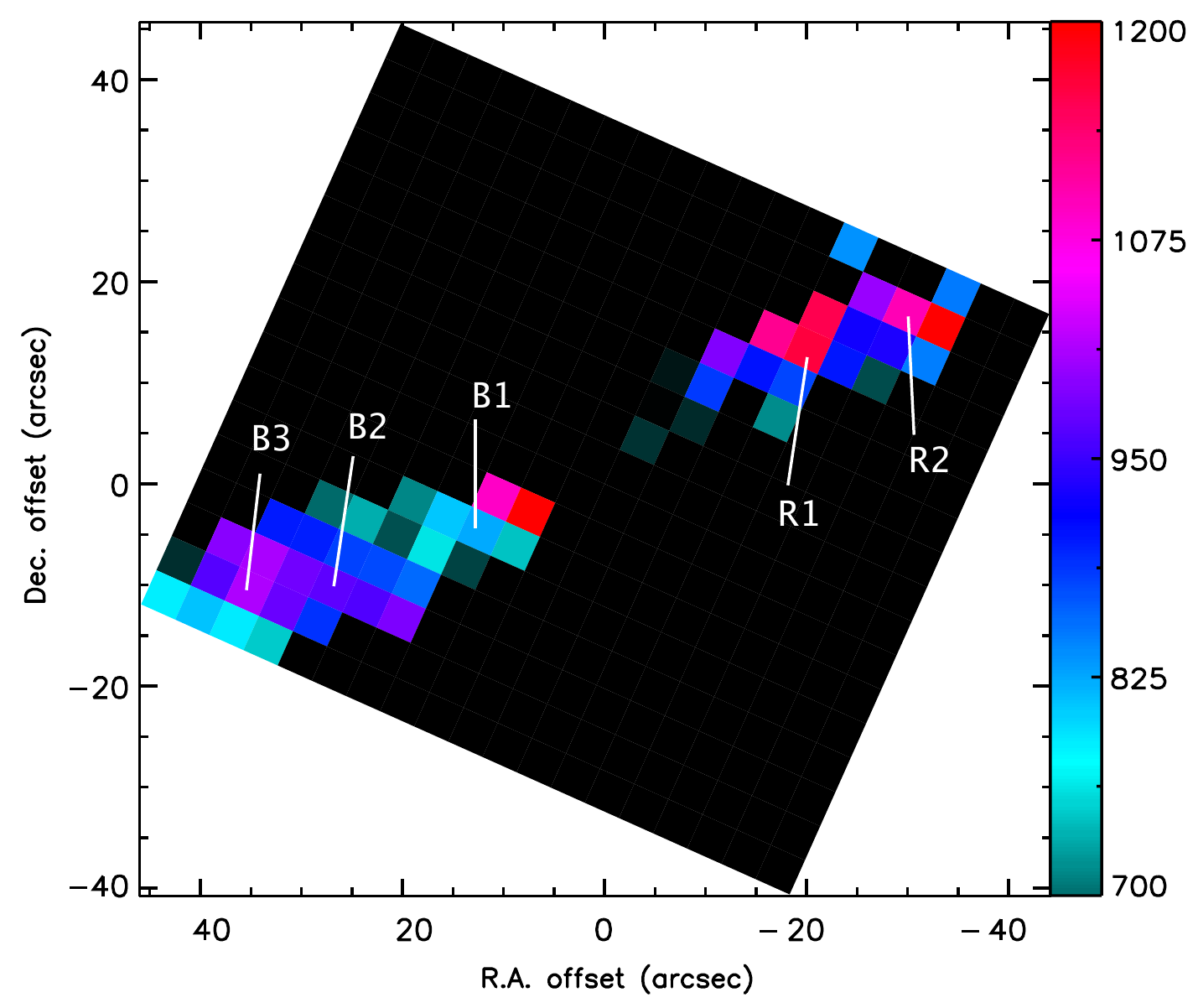}
\caption{Excitation temperature map of the "warm" H$_2$ component deduced from 
the SL module map (3.5$\arcsec$ sampling).  Temperature is derived by a linear fit to the dereddened columns from the v=0-0 S(2) to S(7) lines; Peaks of S(5) emission from Fig. \ref{fig1} are indicated. Pixels with less than 3 detected H$_2$ transitions are in black.}
\label{fig5}
\end{figure}

The results over the whole HH211 region are summarized  
in Fig. \ref{fig5} in the form of an excitation temperature map. Common morphological characteristics between the excitation temperature map and the H$_2$ S(2)-S(7) emissions are evident; peaks in excitation temperature tend to coincide with peaks of emission. At  those peaks, temperature reaches values up to 1200 K, while in  the more diffuse outflow regions the temperature is as low as 700 K. In Table \ref{table1} we report the derived excitation temperatures with this method for the peaks of the S(5) emission along the blue and the red lobes of the outflow. 

In addition to the "warm"  H$_2$ component at $\simeq$ 1100~K indicated by the S(2)-S(7) lines, there is also evidence for a "cool" component from the S(0) and S(1) lines detected in the LL module. This is illustrated in the central panel of Figure \ref{fig4}, which present an EXD of the R1+R2 region in  the LL 10.5$\arcsec$ pixel, including the S(0) and S(1) datapoints.  For consistency, the fluxes of S(2) to S(7) lines from the SL module were re-extracted at the coarser LL module sampling. Due to the extended diffuse emission, the S(0) line flux was obtained after subtracting the ambient contribution estimated in adjacent off-outflow regions.

While the resampled S(3) - S(7) emission keeps the same "warm" 
temperature $\sim$ 1100K, the data points for the S(0)-S(2) transitions  show a steeper slope which is evidence for a  "cool" component at T$\sim$ 300~K not probed by the higher \textit{J} transitions. Another interesting result is that  column densities obtained from the S(2)-S(7) lines in the LL pixel are smaller than in the SL pixel of 3.5$\arcsec$, indicating that the warm component does not fill the LL pixel. 

A similarly resampled EXD towards the LL1 peak also reveals a cool component at 300~K,  in addition to the warm component traced by the $J>2$ lines. (see right-hand panel of Figure \ref{fig4}). Such resampling of the SL data was not possible at the LL peak B2+B3, as part of the corresponding LL pixel falls beyond the SL map coverage.
This double temperature structure differs from a similar Spitzer analysis of the L1448 outflow where all v=0-0 lines from S(0) to S(7) are interpreted with a dominant, single temperature component at 600--900~K. 

\subsubsection{Column Densities and Mass Fluxes}
\label{sec:4.1.2}

The total H$_2$  column density of the "cool" and "warm"  components  can be measured from the intersection of the linear fit to the data points in the EXD with the $N/g$ axis, multiplied by the partition function $Z(T)=\sum{g_i exp(-E_i/kT)}$. The later is estimated for the given temperatures, summing up for the first 35 energy levels of H$_2$.  Derived column densities for the two temperature components are reported in Tables \ref{table1} and \ref{table1.1} and range from $\sim 10^{19}$ for the warm component (averaged over the SL pixel size)
to $8 \times 10^{19}$ cm$^{-2}$ for the cool component (averaged over the LL pixel size).

\begin{table*}
\centering
\caption{Physical properties of the warm H$_2$ component from H$_2$ v=0--0 S(2) - S(7) lines extracted at the SL (3.5$\arcsec$) scale}
\label{table1}
\begin{tabular}{cccccc}
\hline \hline\\[-5pt]
Position & Offsets & H$_2$ 0-0 S(5)& T & N(H$_2$)   & \.{M}$^a$ \\
		& (\arcsec) &	(10$^{-11}$ W cm$^{-2}$ sr$^{-1}$)		& (K)  & 	( 10$^{19}$ cm $^{-2}$) & (10$^{-7}$ M$_{\odot}$\,yr$^{-1}$)\\
\hline\\[-5pt]
R1 &[-20.1, 13.3] & 3.3 &1100$\pm$70 & 1.15  & 1.1\\
R2 &[-29.9, 17.0] & 2.3 &1050$\pm$75 & 1.3 & 1.2\\
B1 &[12.2, -4.7] & 1.3 &860$\pm$40 & 1.1 & 1.1\\
B2 &[26.9, -10.3] & 2.0 &950$\pm$70 & 1.5 & 1.4\\
B3 &[35.1, -10.6] & 1.9 &980$\pm$75 & 1.4 & 1.4\\
\hline\\[-5pt]
\end{tabular} 
\\
~$^a$ \.{M} is proportional to ($V_{jet}$/100 km s$^{-1}$)$\times$(3.5$\arcsec$/$l_t$) and is not
corrected for postshock compression.\\

\end{table*}

\begin{table*}
\centering
\caption{Physical properties of the cool H$_2$ component  from H$_2$ v=0--0 S(0) - S(2) lines extracted at the LL (10.5$\arcsec$) scale}
\label{table1.1}
\begin{tabular}{cccccc}
\hline \hline\\[-5pt]
Position & Offsets & H$_2$ 0-0 S(1)& T & N(H$_2$)   & \.{M}$^a$ \\
		& (\arcsec) &	(10$^{-12}$ W cm$^{-2}$ sr$^{-1}$)		& (K)  & 	( 10$^{19}$ cm $^{-2}$) & (10$^{-6}$ M$_{\odot}$\,yr$^{-1}$)\\
\hline\\[-5pt]

R1+R2 & [-19.3, 15.1] & 1.4 &330$\pm$50 & 7.5  & 2.0 \\
LL1 & [6.3, 2.0] & 1.3 &300$\pm$45 & 7.8 &  2.8\\

\hline\\[-5pt]
\end{tabular} 
\\
~$^a$ \.{M} is proportional to ($V_{jet}$/100 km s$^{-1}$)$\times$(10.5$\arcsec$/$l_t$) and is not
corrected for postshock compression.\\
\end{table*}

As already mentioned in the previous section, the inner peak of H$_2$ emission at B1/LL1 coincides with peaks of the underlying CO and SiO jet; it is therefore instructive to compare
H$_2$ column densities with those deduced from CO, taking into account the differing beam sizes,
to investigate possible physical connections between the different tracers.

\citet{Lee} estimate an H$_2$ column towards the inner CO jet  of 
$4-8 \times$10$^{20}$ cm$^{-2}$ in their 1$\arcsec$ beam, 
assuming an excitation temperature of 100~K and 
a standard CO/H$_2$ abundance ratio of $8.5 \times 10^{-5}$ (ie. a fully molecular gas). 
If the H$_2$ mid-IR emission arises from the same narrow jet, and is relatively uniform along the jet axis,  our derived column densities have to be corrected for the different beam sizes across the jet
by a factor = pixel size / 1$\arcsec$. This increases the H$_2$ columns 
to $\sim 3.5 \times 10^{19}$ cm$^{-2}$  for the "warm" component, and to
$\sim 8 \times 10^{20}$ cm$^{-2}$ for the "cool" component.  The latter value is in excellent agreement with the CO-derived one. This coincidence
 suggests that the "cool" H$_2$ gas detected by Spitzer towards LL1 could be tracing the same material as the CO jet, which would then be indeed {\it mostly molecular}. 
Note also that the temperature 
of 300~K of the cool H$_2$ matches that inferred from SiO line ratios in the jet \citep{Hirano}. 

In contrast, the warm H$_2$ component column density corrected for beam dilution across the jet
remains 10 times smaller than that of the CO jet. This may indicate that the warm H$_2$ emission close to the source arises only from a lower-density outer envelope around the CO jet. Alternatively, the warm H$_2$ component could trace thin shock-heated zones within the CO jet, thus filling only a fraction of the 3.5$\arcsec$ SL pixel length along the jet and suffering further beam dilution. We investigate the shock hypothesis in more detail in the next section (\S \ref{sec42}). 

We may also determine a rough estimate of the one-sided mass flux of the cool and warm H$_2$ components, assuming a uniform laminar flow across the corresponding pixel, using the relationship from \citet{Dionatos}:
\begin{equation}
\label{eq:1.1}
\dot{M} =2  \mu m_{H}\times\langle N(H_2)A\rangle\times(V_t/l_t) 
\end{equation}   
where $\mu$ =1.4 is the mean weight per gram of hydrogen, $m_H$ is the proton mass, $N(H_2)$ the column density averaged over the area $A$ of the pixel, \textit{l$_t$} is 
the projected emitting length along the flow direction (assumed in Tables 1,2 to be equal to the pixel length), and \textit{V$_t$}  the tangential flow speed. 

The value of \textit{V$_t$} is quite uncertain.  Taking into account radial velocities in the range 5 - 20 km s$^{-1}$ measured for the near-IR H$_2$ knots in the study of \citet{Salas}, and assuming an inclination angle between 5$^o$ and 10$^o$ (see \S \ref{sec1}), 
tangential velocities may vary over 30 - 230 km s$^{-1}$, introducing an absolute uncertainty of a factor 3 either way in the inferred $\dot{M}$ values. 
In the following, we adopt V$_t$  = 100 km s$^{-1}$ for ease of comparison with earlier work,
bearing in mind that $\dot{M}$ values scale proportionally to the assumed velocity. 
In column 6 of Tables \ref{table1} and \ref{table1.1} we report the resulting $\dot{M}$ for the peaks of emission along the outflow, for both temperature components. 

For the cool H$_2$ gas component, the laminar mass-flux is $\sim 20 - 28 \times 10^{-7}$ M$_{\odot}$ yr$^{-1}$. 
 For comparison, the one-sided jet mass-loss rate based on CO(3-2) emission obtained by \citet{Lee}  assuming the same velocity of  100 km s$^{-1}$
 and a compression factor of 3  is $3.5-7 \times 10^{-7}$ M$_{\odot}$ yr$^{-1}$. Adjusting their estimations for Helium, and uncorrecting for compression, the laminar  mass flux of the CO jet rises to $15 - 30 \times 10^{-7}$ M$_{\odot}$ yr$^{-1}$; this is in excellent agreement with the cool H$_2$ laminar mass flux estimated here, again consistent with both tracing the same fully molecular gas. 

The warm H$_2$ laminar mass-flux is 15-30 times lower at  $\sim 10^{-7} M_{\odot}$ yr$^{-1}$
$\times (V_t /100 {\rm km~s}^{-1})$.  Similar mass flux estimations were obtained from the pure rotational H$_2$ lines along the outflow of L1448 \citep[$0.5 - 1.4 \times 10^{-7}$M$_{\odot}$ yr$^{-1}$]{Dionatos}, where similar H$_2$ columns were measured\footnote{Note that a typographical error was introduced in Table 2 of \citet{Dionatos}; the values of $N(H_2)$ are $\sim 7 \times 10^{18}$ cm$^{-2}$  at CS and $\sim 1.5-3.8 \times 10^{19}$ cm$^{-2}$  for the other positions.}.
Note however that if the warm H$_2$ does not trace a laminar flow as assumed, but arises in a single shock within the pixel, then the 
relevant emitting length $l_t$ along the flow will be generally narrower than the SL pixel size, and the true mass flux will be higher than listed in Table~\ref{table1}. Mass-fluxes in the warm component in the shock hypothesis are presented in the next section. 

\subsection{Shock models for the warm H$_2$ component \label{sec42}}

As pointed out in \S \ref{sec41}, peaks of warm (\textit{J$\geq$2}) H$_2$ emission coincide with peaks of temperature (Fig. \ref{fig5}) and possibly correspond to regions of shocked gas. In order to constrain the shock conditions in these regions, we employ the existing shock model grid described in the work of \citet{Kristensen} which is based on the MHD-VODE multi-fluid steady shock code of \citet{Flower}. The grid includes both continuous (C) and jump (J) type shocks and predicts the H$_2$ lines brightness for various values of pre-shock density (\textit{n}$_H^{ini}$), shock velocity (\textit{V}$_s$), initial OTP ratio ($otp^{ini}$) and transverse magnetic field density (\textit{$\varphi$}). The ranges of parameters investigated here are as follows:

\begin{itemize}
\item $n_H^{ini}$ = 10$^4$ to 10$^7$ cm $^{-3}$  by alternating factors of 5 and 2. 
\item $V_s$ = 10 - 50 km s$^{-1}$ with a step of 1 km s$^{-1}$
\item $otp^{ini}$ = 0.01, 1, 2, 3
\item $\varphi (\mu G) = b \times \sqrt{n_H \, (cm^{-3})}$ where we fix \textit{b}=1 for C and \textit{b}=0 for J-type shocks. 
\end{itemize}

The model is one-dimensional, considers 9 elements (H, He, C, O, N, S, Si, Mg and Fe) and 136 species connected with 1040 reactions, and takes into account grain sputtering and erosion, as well as a variety of cooling and heating processes. Initial chemical abundances are computed assuming no UV field and an H$_2$ cosmic ray ionization rate of $5 \times 10^{-17}$ s$^{-1}$, yielding an ionization fraction $\sim 7\times 10^{-8} (n_H^{ini}/10^4 {\rm cm~s^{-1}})^{-0.5}$.
The H$_2$ level populations and line brightnesses are integrated along the postshock cooling zone down to a temperature of 50~K, which is reached in a cooling time  $\sim$ 100 ($10^5$ cm $^{-3}$/ $n_H$) yrs in C-shocks (much shorter in J shocks).  For more information on both the model and the grid data, the interested reader is referred to the articles of  \citet{Flower} and \citet{Kristensen}.  

In the following, we have employed a $\chi^2$ fitting method to optimally reproduce the observed H$_2$ emission selecting the best matching C or J-type steady shock models from the grid. The $\chi^2$ - fitting is performed with and without considering the S(2) transition, since the latter may be dominated by the cool component (see Fig.\ref{fig4}).  

\subsubsection{C-Shocks}

As a general trend we found that steady C-shock models are able to reproduce the observed S(2)-S(7) brightnesses only if we assume a small beam filling factor. For this reason we  have performed our $\chi^2$ fits on relative line brightnesses, scaled to the H$_2$ S(5) line; for the best fitting models we have then calculated a beam filling factor as the average ratio of the observed over model line brightness.  

\begin{figure}
\includegraphics[scale=0.4]{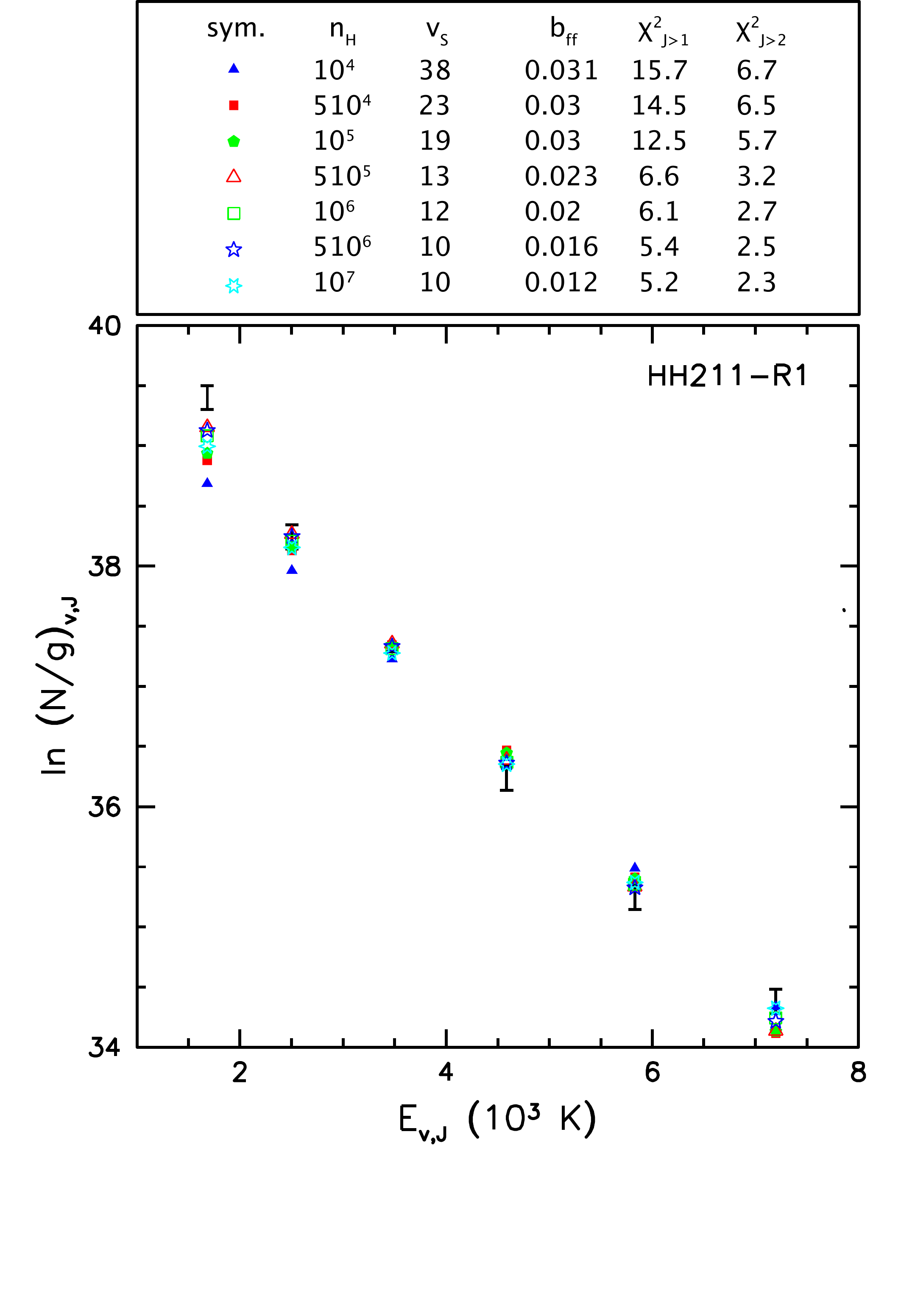}
\caption{Steady 1D C-shock model fits to the excitation diagram of the S(2)--S(7) lines at the peak of emission R1 in a 3.5$\arcsec$ pixel. Observed points (filled circles with 10\% error bars) are displayed along with  best fitting $\chi^2$ models for pre-shock densities between  10$^4$ and 10$^7$ cm$^{-3}$. The top panel lists the corresponding shock velocities and beam filling factors, along with the reduced $\chi^2$ fit values to the S(2) - S(7) (J$>$1) and to the S(3) - S(7) (J$>2$) H$_2$ transitions. A strong degeneracy among models is found (see text).}
\label{fig7}
\end{figure}

In Figure \ref{fig7} we plot the EXDs of the best fitting $\chi^2$ models  for each preshock density, scaled by the corresponding beam filling factor, along with the observed data (filled circles) for the peak of emission R1; note that beam filling factors decrease for increasing values of density. 
There is a strong degeneracy among the models, with a very small spread in column densities for
$n_H^{ini}$ in the range between 5 10$^5$ and 10$^7$ cm$^{-3}$ and velocities in the range 10-13 km s$^{-1}$, however models for lower values of n$_H \leq$ 10$^5$ and velocities up to 30 km s$^{-1}$ cannot be excluded; each one of these models can account fairly well for the observed emissions. 

\begin{figure*}[!t]
\includegraphics[scale=0.3]{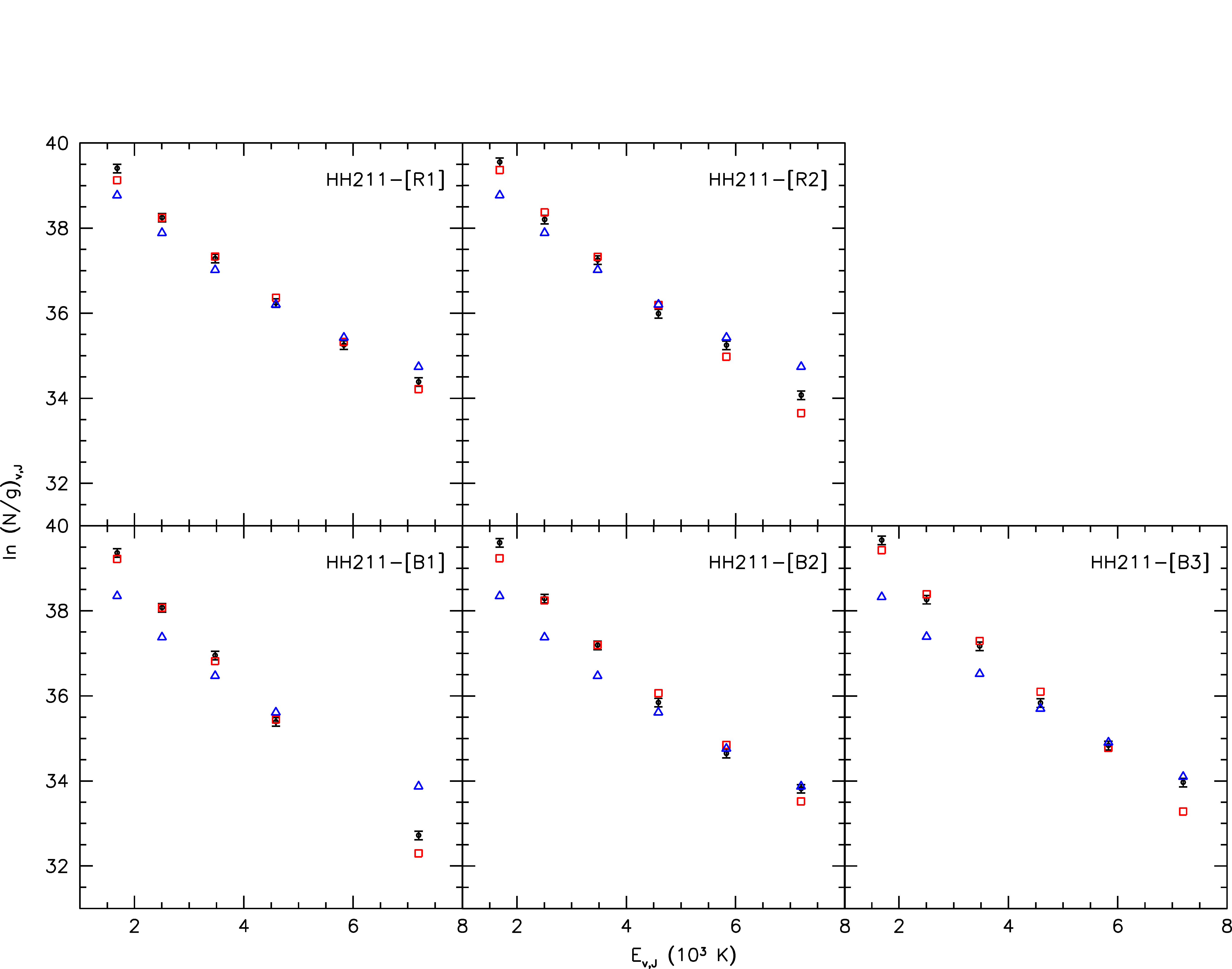}
\caption{Excitation diagrams for all emission peaks mapped with the SL module; observed points (filled circles with error bars), are displayed along with best $\chi^2$ fitting C (red squares) and J (blue triangles) shock models. C- shock models can account fairly well for the observed emissions considering a small beam filling factor, whereas J-shock models cannot reproduce well the slopes of the observed data points, except marginally in R1 and R2.}
\label{fig8}
\end{figure*}

Figure \ref{fig8} presents EXD of the best matching C--type shocks (red squares, always scaled by the corresponding beam filling factor) along with the observed data points (filled circles), for the peaks of emission as observed along the outflow. Input parameters for models are reported in Table \ref{table2}. All  C-shock models predict moderate shock velocities (10-15 km s$^{-1}$) and an initial ortho to para ratio at the high temperature LTE value of 3.

Steady C-shock models reproduce the data only if we consider low beam filling factors \textit{b$_{ff}$}  $\sim$ 0.01-0.04, or in other words a shock area \textit{A$_{shock}$ = A$_{pixel} \times$ b$_{ff}$}, corresponding to a typical size of $\sqrt{A_{shock}}\sim$0.3$\arcsec$ - 0.7$\arcsec$. These dimensions are comparable to the reported jet width $<$1$\arcsec$ in the interferometric CO and SiO maps of \citet{Lee}. High post-shock densities in the same range as the best fitting C-shock models (5 10$^5$ - 10$^7$ cm$^{-3}$) have been retrieved in the jet by SiO emission studies \citep{Nisini1, Hirano, Lee}, supporting our modeling results. The shock models predict thin H$_2$ post-shock cooling zones of 0.015\arcsec - 0.25\arcsec, much smaller than the SL pixel size of 3.5$\arcsec$, explaining the small warm column densities derived in the previous section. The cooling zone is also smaller than the shock surface size inferred from the filling factor, as required for the 1-D planar shock approximation to be locally valid. 

If the warm H$_2$ emission indeed traces low-velocity internal shocks within a fast jet beam, we may infer the jet mass-flux from the pre-shock density and the shock area as: 
\begin{equation}
\dot{M}_{j,shock} = \mu m_{H}\times  b_{ff} A_{pixel} \times n_{H}^{init} \times V_{jet}. 
\label{eq:mdotshock}
\end{equation}
Considering a pre-shock density of $5 \times 10^{5}$ cm$^{-3}$ with a beam filling factor $b_{ff}$ of 0.023, and a jet speed of 100 km s$^{-1}$, we obtain  for the warm H$_2$
$\dot{M}_{j,shock} = 7 \times 10^{-7}$ M$_\odot$ yr$^{-1}$. This is 7 times higher than the value obtained in Section 4.1.2 assuming a laminar flow filling the SL pixel.  
The mass-flux entering the shock is only a factor 3-4 below the  laminar mass-flux for the cool H$_2$ (and CO). Such a discrepancy is  expected in a time-variable flow undergoing internal shocks \citep{hart94}. Therefore, the "warm" H$_2$ component seems compatible with localized heating in slow internal C-shocks within the molecular jet beam, and may not trace a separate lower density flow. Note that pre-shock densities higher than 5 $\times$ 10$^6$ cm$^{-3}$ seem to be excluded, as the pre-shock mass-flux rate from Equ.~\ref{eq:mdotshock} would become comparable to the accretion rate onto the HH211-mm driving source \citep[8 $\times$ 10$^{-6}$ M$_\odot$ yr$^{-1}$][]{Lee}. 

However, SiO is observed only within $\sim$15$\arcsec$ from the source, therefore pre-shock densities of $5 \times 10^{5}$ cm$^{-3}$ might apply only to the B1 region. In addition, in our maps the H$_2$ emission remains unresolved only close to the driving source. Further out at the bow-shocks, the emission appears slightly resolved at the 3.5$\arcsec$ scale. A small $b_{ff} <0.04$ would therefore imply that the bow-shock surface is highly patchy. 
This is a reasonable possibility, since HST observations in several outflows have shown that what
appears as an individual bow-shock at low resolution is in fact resolved in many 
sub-structures of sub-arcsec scale (e.g. HH2 in Orion, \citet{bally}). In HH211, SMA observations, performed with a resolution of $\sim$ 1$\arcsec$ \citep{Lee}, show indeed
that the CO emission is highly clumped, both in the high velocity jet and 
along the low velocity cavity ahead of the bow-shocks.

An alternative explanation for the small beam filling factor in the outer HH211 bow-shocks might be that the pre-shock density drops to $\le$ 10$^4$ cm$^{-3}$, so that the cooling length becomes comparable to the 3.5$\arcsec$ pixel size; the hypothesis underlining our model comparison, that the whole cooling zone fits inside the pixel, would then become invalid. Indeed, \citet{Oconnell} derived a low pre-shock density in the R1 and R2 regions by fitting a 3D C-type bow-shocks to the near-IR ro-vibrational H$_2$ lines. However, no predictions for the mid-IR H$_2$ emissions were provided. Extension to 3D bow-shock geometry of the model applied here \citep{gustafsson} will be needed to perform more reliable modeling of the outer bowshocks of HH211 than possible in the present analysis. 

A third possibility might be that the preshock gas ahead of the leading bowshocks is more ionized than we have assumed, due to the presence of a UV flux from nearby dissociative shocks such as that observed by \citet{Tappe} in the blue-shifted apex. This
would increase the ion-coupling in the C-shock, reducing its width and column density, and yielding a larger filling factor. Indeed, \citet{Oconnell} find that a 
high pre-shock ionization fraction of 10$^{-5}$ and a high atomic hydrogen fraction of 20$\%$
provide a good fit to the near-IR lines in the redshifted bow. 
However, given the already high degeneracy of our model fits, and the lack of information on
shock geometry and thickness in low angular resolution Spitzer maps,
introducing the UV field as an additional free parameter is not feasible in practice.
Additional constraints from other line tracers or from higher resolution mid-IR maps (eg. with JWST)
will be required to further explore this alternative.

\begin{table*}
\centering
\caption{Parameters of shock models best fitting the S(3) to S(7) lines}
\label{table2}
\begin{tabular}{ccccc|ccc}
\hline \hline\\[-5pt]
Peak & \multicolumn{4}{c}{C-shock (b=1)}  & \multicolumn{3}{c}{J-shock (b = 0, b$_{ff}$=1)}\\ 
     &$n_H^{ini}$(cm$^{-3}$) & V$_s$(km s $^{-1}$) & opr$_{ini}$ & b$_{ff}$ & $n_H^{ini}$(cm$^{-3}$) & V$_s$(km s$^{-1}$) & opr$_{ini}$ \\ 
\hline\\[-5pt]
R1 &  5 10$^5$ - 10$^7$ & 10 - 13 & 3 & 0.01 - 0.02 &  5 10$^4$ & 14 & 3\\
R2 &  5 10$^5$ - 10$^7$ & 10 - 11  & 3 & 0.02 - 0.04 &  5 10$^4$ & 14 & 3\\
B1 &  5 10$^5$ - 10$^7$ &  10  & 3 & 0.02 - 0.03 &   10$^4$ & 10 & 3\\
B2 &  5 10$^5$ - 10$^7$ & 10 - 11  & 3 & 0.01 - 0.03 &   10$^4$ & 10 & 3\\
B3 &  5 10$^5$ - 10$^7$ & 10 - 11  & 3 & 0.02 - 0.03 &   10$^4$ & 11 & 3\\
\hline\\[-5pt]
%
\\
\end{tabular}
\end{table*}

\subsubsection{J-Shocks and non-steady C-shocks}

In the case of J-shocks, absolute brightness can be fairly well matched by models and beam filling factors $< 1$ were not necessary to be considered. The predicted pre-shock densities  for J-shock models $\sim 1- 5\times 10^4$ cm$^{-3}$ are lower than for the best fitting C-shock ones. As a general trend though, the best fitting J-shock models, presented with blue triangles in Fig. \ref{fig8} and listed in Table~\ref{table3}, cannot reproduce  at the same time the {\it slope} of the  observed data points, except for the red-shifted bowshock (R1 and R2). 

An analogous behavior has been noted in the work of \citet{Giannini}, where models produced by the same code of \citet{Flower}  are employed to study the conditions of the HH54 outflow. Considering only the rotational H$_2$ emission in reported excitation diagrams, \citet{Giannini} find that steady C-type shocks overpredict the observed v=0 column densities, whereas J-type shocks can fall much closer but cannot fit at the same time the observed slope. In the case of HH54, it is concluded that a non-steady, truncated C-shock with an embedded J-type front (C+J shock) optimally reproduces the observed H$_2$ emission. The same conclusion was reached also in the case of  L1157  \citep{Gusdorf2}. Indeed, the overall effect as reflected in excitation diagrams is to lower the column densities of low excitation energy levels (truncated C-precursor) and increase the column densities of the rovibrational energy levels of H$_2$, excited in the J-type front \citep{FlowerCJ}. In our case, such a model might potentially fit our mid-IR data in the outer bowshocks without the need to consider a beam filling factor $< 1$. A more detailed comparison with unsteady C+J-shock models would be needed to test this possibility, however such models were not included in our adopted grid as the age of the shock would have, again, added an extra free parameter to already degenerate fits.  The shock ages would have to be much shorter than the outflow dynamical age
of about 1000 yrs in order to reduce by a factor 50 the warm H$_2$ column (typically a shock age of 1/50 of the $H_2$ cooling time, ie. $\sim$ 2 ($10^5$ cm $^{-3}$/ $n_H$) yrs).

\begin{table*}[!t]
\centering
\caption{Physical properties of the atomic/ionic line component}
\label{table3}
\begin{tabular}{ccccccc}
\hline \hline\\[-5pt]
Position & Offsets & [SI] & [FeII]/[SI] & [SiII]/[FeII] &\.{M}([SI])$^{a,b}$ & $n_H^{b,c}$  \\ 
    &(\arcsec) &(10$^{-13}$ Watt cm$^{-2}$ sr$^{-1}$) &  & &(10$^{-7}$ M$_{\odot}$\,yr$^{-1}$) & (10$^5$ cm$^{-3}$) \\
\hline\\[-5pt]
R1+R2 &[-19.3, 15.1]& 5.34 $\pm$ 0.13 & 0.92 $\pm$ 0.04 & 2.18 $\pm$ 0.51  &0.4 - 3.4 & 0.21 - 1.61\\
LL1      &[-6.5, 8.6]    & 6.24 $\pm$ 0.16 & 0.60 $\pm$ 0.02 & 1.77 $\pm$ 0.82   &0.4 - 4.1 &0.24 - 1.88 \\
LL2      &[6.3, 2.0]     & 7.49 $\pm$ 0.19 & 0.35 $\pm$ 0.01 & 1.95 $\pm$ 0.93 & 0.7 - 4.9 & 0.29 - 2.91\\
B2+B3 &[31.9, -11.0] & 16.1 $\pm$ 0.41 & 0.37 $\pm$ 0.01 & 2.14 $\pm$ 0.42  & 1.3 - 10.5& 0.62 - 4.9 \\
\hline\\[-5pt]
\end{tabular}
\\
~$^a$ \.{M} is proportional to ($V_{jet}$/100 km s$^{-1}$)$\times$(10.5$\arcsec$/$l_t$) and is not
corrected for postshock compression.\\
~$^b$ values calculated for collisions with electrons at T =  3000~K (lower value) and T = 700~K (upper value); including collisions with hydrogen for n(H)=10$^4$ cm$^{-3}$ would decrease the minimum and maximum values by a factor of $\sim$ 2 and 4 respectively.\\
~$^c$average proton density assuming an emitting volume of 10.5$\arcsec \times$ 1$\arcsec \times$ 1$\arcsec$
\end{table*}

\subsection{Atomic emission\label{sec5}}

As pointed out in section \S \ref{sec3}, all atomic/ionic emission falls within the LL IRS module and consequently these maps are of inferior spatial resolution compared to the SL ones. As atomic emission is found to be enhanced at the bow shocks but remains significant also close to the driving source, we have focused our analysis all along the jet axis including the intermediate points LL1 and LL2 (see Fig. \ref{fig3}) between the bow shocks, where the [FeII] 26$\mu$m  line is detected. For the analysis of the ionic emission we have followed the techniques employed by \citet{Dionatos}, and the reader is referred to this article for a detailed description.

\subsubsection{Electron density}

\begin{figure}
\includegraphics[width=9cm]{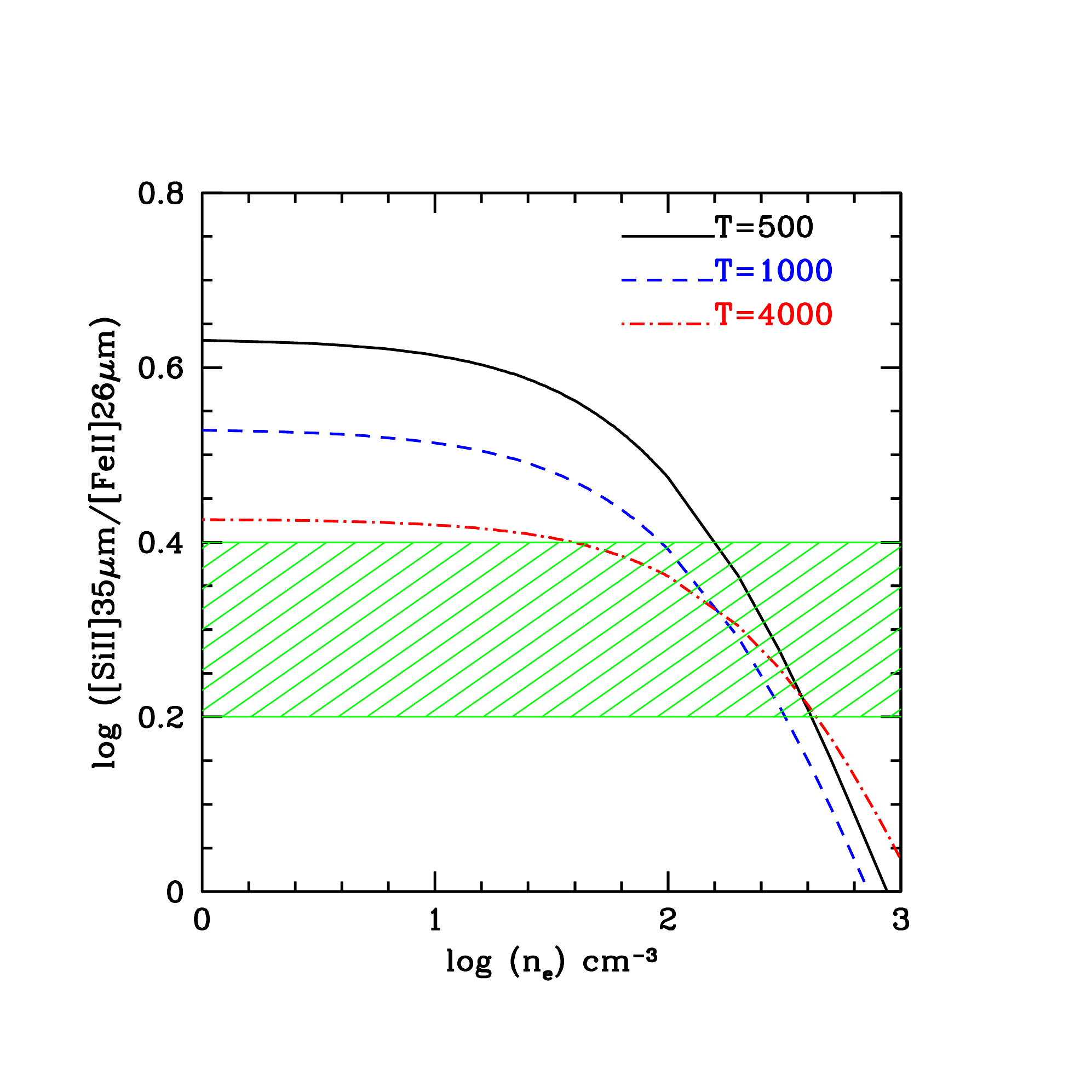}
\caption{Diagnostic diagram of the [SiII]35$\mu$m/[FeII26]$\mu$m ratio versus electron density, for temperatures of 500K, 1000K and 4000K (solid and dashed lines) and assuming a solar Si$^+$/Fe$^+$ abundance ratio. The hatched area represent the observed ratio for the north and south bow-shock regions, from which electron densities between $\sim$ 100 and 400 cm$^{-3}$ are found .}
\label{fig9}
\end{figure}

As a first step, the observed emissions of [SiII] and [FeII] are employed in order to constrain the electron density. In Figure \ref{fig9}  the ratio of [SiII]34.8$\mu$m over [FeII]26.0$\mu$m as a function of the electron density is presented, for temperatures of 500K, 1000K and 4000K (solid and dashed lines).  A statistical equilibrium model was employed that uses the first 16 levels for [FeII] \citep{Nisini2}, and a two level system was considered for [SiII], with radiative and collisional rates with electrons taken from \citet{Dufton}; a solar abundance ratio of Si$^+$/Fe$^+$ was considered, taking into account the similar ionization potential of the two species as well as the similar  efficiency of sputtering of Si and Fe atoms from olivine grains in C-shocks, according to Figure 2 of \citet{Gusdorf}. 
In the diagram we plot the observed ratio limits at the bow shocks for clarity, but the results for the other points along the jet are similar and are listed in Table \ref{table3}.  Lower limits for the electron density vary between 40 and 150 cm$^{-3}$ for temperatures of 4000K and 500K  respectively, while upper limits converge for the considered temperatures to $\sim$ 400 cm$^{-3}$. Temperature limits lower than a few thousand K are reported in the work of \citet{Tappe} for the blue-shifted bow shock considering a non-LTE analysis of the three [FeII] lines (18 $\mu$m, 26$\mu$m and 35$\mu$m) that are detected with the high resolution IRS modules. Based on this, we
restrict the electron density lower limit to 100 cm$^{-3}$ by assuming a maximum temperature of 3000~K.

\subsubsection{Gas phase depletion\label{sec4.3.2}}

We investigate the depletion onto grains following the analysis of \citet{Dionatos}. We calculate the gas phase abundance of refractory elements  such as iron and silicon by comparing their emissions against the emission of a non-refractory species, in the present case sulphur.  Theoretical emissivities for sulphur were calculated considering a five level statistical equilibrium code for an upper temperature limit of 3000 K assumed from the work of \citep{Tappe} and a lower limit of 700 K as imposed from the H$_2$ analysis. For the excitation of sulphur, we have also calculated emissivities due to collisions with atomic hydrogen, taking into account that in environments with low ionization fraction and high total density such excitation may prove to be significant, despite the fact that collisional de-excitation rates for atomic hydrogen are lower than the electronic ones \citep[$\gamma_{H}/\gamma_{e} \sim 0.02$]{Hollenbach} for the [SI]25$\mu$m line. For these calculations
 we have considered a medium with 
 n(H) = $10^4$ cm $^{-3}$. Excitation of [FeII] and [SiII] by H-collisions was not taken into account because neutral hydrogen collisional rates are much less important for ions (i.e. $\sim$ 10$^3$-10$^4$ weaker than electronic collisional rates), and are not available for all the levels considered in our model.

\begin{figure}
\includegraphics[width=9cm]{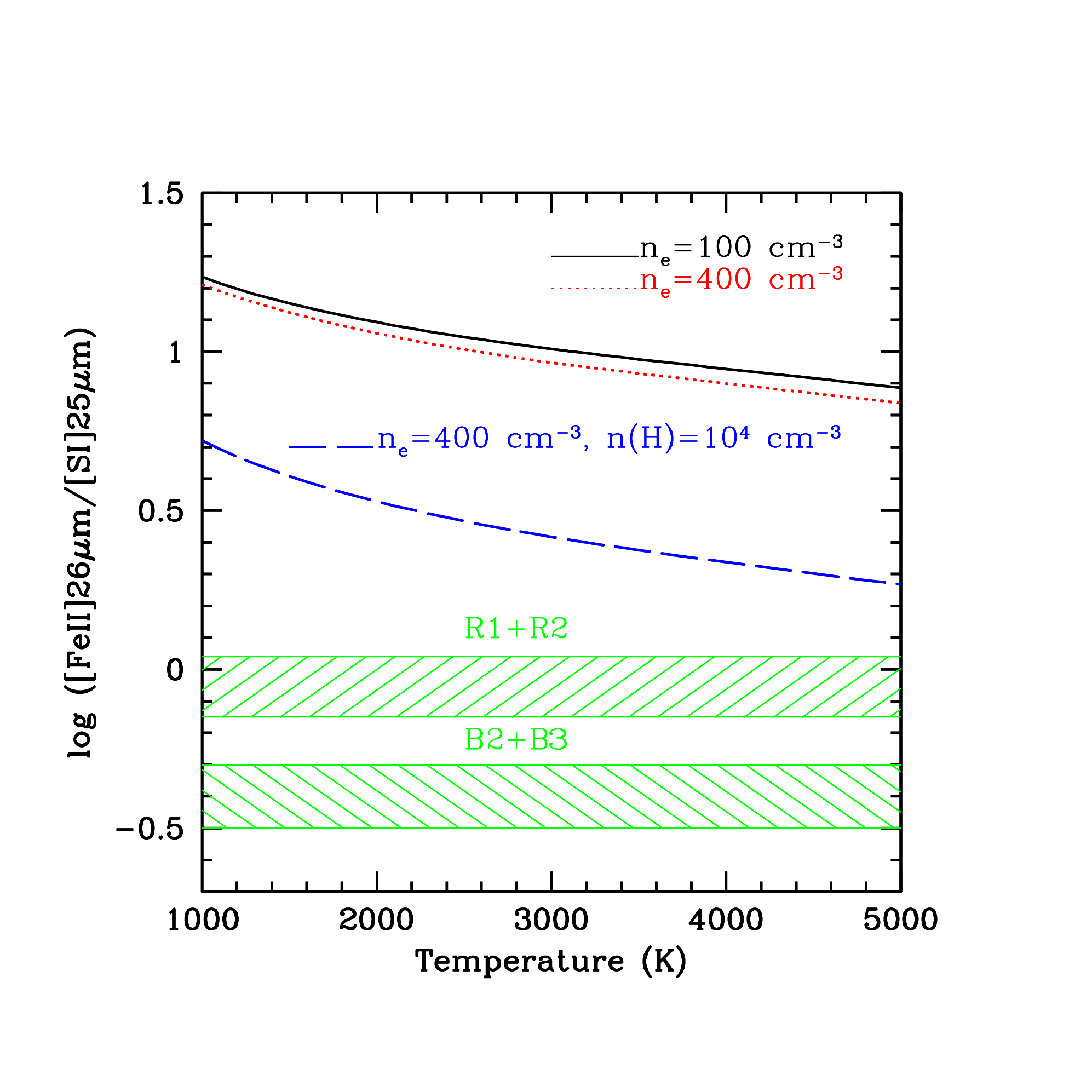}
\caption{Diagnostic diagram of [FeII]26$\mu$m/[SI]25$\mu$m ratio versus temperature
assuming solar relative abundances. Electron densities of 100 and 400 cm$^{-3}$ are considered (full and dotted lines, respectively). The effect of collisions with atomic hydrogen on the [SI] excitation is also shown (long-dashed line, $n(H)=10^4$ cm$^{-3}$).  The observed ratios are always below the predicted values for solar abundances, which is interpreted as iron being depleted onto dust-grains in the considered regions.}
\label{fig10}
\end{figure}

Results of these calculations are illustrated in Fig. \ref{fig10} where we plot the ratio of  [FeII] 26$\mu$m over [SI] 25$\mu$m  against gas temperature for  electron densities of 100 and 400 cm$^{-3}$, assuming that iron is fully ionised (which is the case in most dissociative shocks) and has a solar abundance. The observed ratios for the points R1+R2 and B1+B2 are way below the expected values, giving rise to the hypothesis that iron is heavily depleted onto dust grains in the considered regions, more so if the emitting gas is cooler. The case including atomic hydrogen collisions with n(H) = $10^4$ cm $^{-3}$ is  separately plotted in Fig. \ref{fig10}. The iron depletion is less severe in this case but is still substantial, about a factor 10. 

In Table~\ref{table4} we give the inferred gas phase depletion for temperatures of 700 and 3000 K, both without and with atomic hydrogen excitation of [SI]. The gas phase abundance of  Fe$^+$ is about 2--20\% of the solar value for T= 3000 and 700~K respectively, for all the points under examination. Similar low values were found also in the case of the L1448 jet (5--20$\%$). Identical gas-phase depletions are obtained for Si$^+$, but since we assumed a solar Fe$^+$/Si$^+$  ratio in the first place to derive our $n_e$ values, this is just a self-consistency check. Interestingly, the SiO abundance in the HH211 molecular jet,  estimated as $\sim 10^{-6}$ by \citep{Nisini1, Chandler} corresponds to 3\% of the solar abundance of silicon, therefore the fraction of silicon released in the gas phase seems comparable in the atomic and molecular components of HH211. 

\begin{table}
\centering
\caption{Gas phase abundance of Fe$^+$ and Si$^+$ from line ratios to sulfur}
\label{table4} 
\begin{tabular}{ccccc}
\hline \hline\\[-5pt]
Position: &  R1+R2 & LL1  & LL2 & B2+B3\\
\hline\\[-5pt]
$[{\rm Fe}^+_{gas}]/[{\rm Fe}_{\odot}]^a$  & 4-9 10$^{-2}$ & 3-6 10$^{-2}$ & 2-4 10$^{-2}$ & 2-4 10$^{-2}$\\
$[{\rm Fe}^+_{gas}]/[{\rm Fe}_{\odot}]^b$ & 10-20 10$^{-2}$ & 7-10 10$^{-2}$ & 5-7 10$^{-2}$ & 5-7 10$^{-2}$\\
\hline
$[{\rm Si}^+_{gas}]/[{\rm Si}_{\odot}]^a$ & 4-9 10$^{-2}$ & 2-4 10$^{-2}$ & 1.5-3 10$^{-2}$ & 2-4 10$^{-2}$\\
$[{\rm Si}^+_{gas}]/[{\rm Si}_{\odot}]^b$ & 10-20 10$^{-2}$ & 5-8 10$^{-2}$ & 3-6 10$^{-2}$ & 4-7 10$^{-2}$\\
\hline\\[-5pt]
\end{tabular}
\\
~$^a$ assuming only collisions with electrons ($n_e =100-400$ cm$^{-3}$) and $T_e$=700~K (lower value) and 3000~K (upper value)\\
~$^b$ including excitation of sulfur by atomic hydrogen with $n(H)=10^4$ cm$^{-3}$ \\
\end{table}

\subsubsection{Density and mass flux of the atomic component}

The total number of H nuclei in the emitting volume $V$ can be derived from the line luminosity according to the relation:
\begin{equation}
\label{eq:4}
n_{H}\,V=L(line) \left(h\nu A_i f_i \left[\frac{X}{H}\right] \right )^{-1},
\end{equation}
where $A_i$ and $f_i$ are the spontaneous radiative decay rate and fractional population of the upper level,  and [X/H] is the gas phase abundance of the atom/ion under consideration with respect to H nuclei. Line luminosities were integrated over the LL pixel size area, and the fraction of atoms at the upper level $f_i$ was calculated for $n_e$ between 100 and 400 cm$^{-3}$ and T= 700-3000K as imposed from the previous analysis. 

This relation was applied to [SI], the only non-refractory species detected, assuming that all sulfur is in neutral form and employing its solar abundance \citep{Asplund}. Such calculations for refractory species would be redundant, as their gas phase abundances from Table~\ref{table4}  were determined from their intensity ratio to [SI]. 

Values of $n_H$ at individual positions are listed in column 7 of Table~\ref{table3} assuming
an emitting volume in the LL pixel of 10.5$\arcsec \times$ 1$\arcsec \times$ 1$\arcsec$ (uniform narrow jet). The inferred $n_H$ depends strongly on the adopted excitation conditions:  it is about $0.2-0.6 \times 10^5$ cm$^{-3}$  for T=3000~K, $n_e = 100$ cm$^{-3}$, and 8 times higher for T=700~K, $n_e = 400$ cm$^{-3}$. The ionisation fraction is then
$1.6-5 \times 10^{-3}$ at 3000~K, or twice smaller at 700~K. Such an ionization level seems to favor the upper temperature, and therefore the lower density range. If the fraction of H atoms is important with $n(H) \simeq 10^4$ cm$^{-3}$, the $n_H$ values would be further reduced by a factor 2, down to a few $10^4$ cm$^{-3}$. In the case again, that the emission comes from a smaller volume, eg. a small 1$\arcsec$ knot as observed in the near-IR [FeII]1.64$\mu$m line, the density would rise up to $10^5$ cm$^{-3}$.

The mass flux in the atomic/ionic component assuming a laminar flow can also be derived applying the relationship given in \citet{Nisini3} as applied to the mid-IR lines in \citet{Dionatos}:
\begin{equation}
\dot{M} = \mu m_{H}\times(n_{H}\,V)\times(V_t/l_t),
\end{equation}
where $\mu$=1.4 is the mean weight per H nucleus, $m_H$ is the proton mass, $n_H\,V$  is the total number of protons in the emitting region as given by Eq. \ref{eq:4}, $l_t$ is the projected emitting length along the flow and $V_t$ the tangential flow velocity. 

Derived mass flux values at the various emission peaks are listed in column 6 of Table \ref{table3} for a tangential speed of 100 km s$^{-1}$ and $l_t =10.5\arcsec$ (LL pixel size). Like 
$n_H V$, they again depend strongly on the adopted excitation conditions. From the ionization fraction considerations made above, we favor the high temperature, smaller $\dot{M}$ values
of $\sim 0.4 - 1 \times 10^{-7}$ M$_{\odot}$ yr$^{-1}$. These are 20--50 times smaller than the cool H$_2$ jet mass-flux, assuming the same flow speed of 100 km s$^{-1}$. 

However, we stress again that $\dot{M}$ estimates assuming a laminar jet flow may be in error if the [SI] emission arises from shocks, which is likely given the ionization fraction inferred above. This is difficult to quantify without appropriate shock models.

The atomic/ionic emission appears to require higher excitation shocks than those producing the mid-IR  H$_2$ lines.  The low velocity 10--15 km s$^{-1}$ shock models that best fit the warm H$_2$ emission are unable to reproduce also the observed [FeII] and [SiII] line intensities unless the shocks are of J-type and these atoms are essentially undepleted. As discussed in Section \ref{sec4.3.2} there is overwhelming evidence of depletion of these refractory species, so [FeII] and [SiII] probably arise from faster shocks than those dominating the mid-IR H$_2$ lines. It is indeed likely that the mid-IR atomic lines originates from the same dissociative shocks at $v_s \ge 30$ km s$^{-1}$ that give rise to the optical and NIR emission of H$_{\alpha}$, [SII] and [FeII] \citep{Oconnell,caratti,walaw1,walaw2}.

Unfortunately, our shock models do not yet include ionization and dissociation by the shock UV flux, so we cannot explore the relevant excitation range. 
However, we note that the J--shock models of \citet{Hollenbach}, which include UV flux, suggest that the [SI]25$\mu$m brightness is relatively independent of shock speed over the range 30--100 km s$^{-1}$ and is roughly proportional to $n_H^{ini}$. Our measured flux inside the LL 10.5$\arcsec$ pixel would suggest $n_H^{ini} \times b_{ff} \simeq 100 - 1000$ cm$^{-3}$. From Eq. \ref{eq:mdotshock} we would infer a preshock mass-flux of 
$\sim 0.5 - 5 \times 10^{-7}$ M$_{\odot}$ yr$^{-1}$, still smaller by a factor 4--40  than the cool H$_2$ jet mass-flux.

These results would suggest that the atomic component in the HH211 jet does not have enough momentum flux to entrain the CO/SiO/H$_2$ jet, {\it if} their velocities are comparable and the shock compression in the cool H$_2$/CO component does not exceed a factor 3. The molecular jet would then have to trace material ejected from the accretion disk, while  the atomic component would trace a separate ejection, e.g. from hotter more internal regions of the accretion disk. 
On the other hand, a higher compression factor in the cool H$_2$/CO emitting zone, or a faster ionic jet would suffice to remove the discrepancy, therefore a definite conclusion on whether the molecular jet is ejected or entrained cannot be reached from the present data alone. 

\section{Conclusions\label{sec6}}
  
We have carried out Spitzer spectral mapping observations towards the jet driven by the Class 0 source HH211-mm.  Molecular lines (pure rotational H$_2$) as well as fundamental atomic and ionic lines ([SI], [SiII], [FeII]) were detected, and their maps follow the characteristic bipolar outflow pattern as traced by near-IR H$_2$  and CO lines. H$_2$ emission becomes important only 5\arcsec\ away from the driving source while atomic and ionic lines are detected very close to the driving source. In the inner part of the blue-shifted lobe, the H$_2$ emission is spatially coincident with the high velocity jet observed in CO and SiO. 

Analysis of the observed H$_2$ lines reveal two temperature components: "cool" gas at T$\sim$ 300K dominating the mass, and "warm" gas at T$\sim$ 700-1200K of 10 times smaller
average column density. Once corrected for beam dilution, the column density of "cool" H$_2$ towards the CO jet is compatible with hydrogen being mostly molecular. The warmer component traced by the S(2) to S(7) lines is well fitted by C-shock models of high density $\simeq 5 \times 10^5$ and a small shock cross section of 0.5$\arcsec$, compatible with the density and width of the CO jet quoted by \citet{Lee}. The warm H$_2$ emission could then trace thin layers of warm post-shock gas within the CO jet. 

Similarly, high-density C-shock models can also account for the brightness of the H$_2$ mid-IR emission further downwind beyond the CO and SiO jet, but the small shock surface filling factor of 0.01-0.04 is not easy to reconcile with the spatial extension visible in Spitzer maps, unless the bow-shock wings are very clumpy. Lower density shocks of n$_H$ $\leq$ 10$^4$ cm$^{-3}$ where the cooling zone spreads over several pixels and/or higher preshock ionization may need to be considered in these regions, as previously invoked to model near-IR H$_2$ excitation in the red-shifted bow-shock of HH211 \citep{Oconnell}.

The detected fine structure lines mapped very close to the driving source signify the presence of an embedded atomic jet. Line ratio diagnostics indicate a gas-phase depletion of iron and silicon of at least a factor 10, and lower excitation conditions than in optically visible jets.  An atomic jet of similar properties has been also detected by Spitzer in the outflow of L1448-C \citep{Dionatos}. As suggested in that case, the detected atomic gas in the HH211-mm outflow may represent the equivalent for the Forbidden Emission Line (FEL) region observed in more evolved ClassI/II sources. The gas depletion of iron and silicon indicates that dust grains have survived in the atomic flow, ruling out an origin from within the dust sublimation zone close to the protostar. The excitation conditions require faster shocks than those producing the  H$_2$ mid-IR lines.

Estimations of the molecular and atomic mass-flux rates have been performed using both a laminar flow assumption and shock models. The cool H$_2$ mass-flux is comparable to that inferred from CO observations, with  a value uncorrected for compression of $\sim 2 \times 10^{-6}$ M$_{\odot}$ yr$^{-1}$ ($V$/100 km s$^{-1}$); a similar value is found for the warm H$_2$ if it arises from dense C-shocks as suggested by best fit models. The atomic component mass-flux is uncertain by a factor 8 due to its uncertain temperature, but considerations of ionization fraction as well as published dissociative shock models suggest a mass-flux smaller than in the molecular jet. However, the momentum fluxes could become comparable if the atomic jet is faster than the molecular jet, or if the cool H$_2$ and CO suffered shock compression.  Given the uncertainties,  it is still unclear whether the molecular CO jet is tracing ambient gas entrained by the atomic jet, or if it has to be ejected, for example in a molecular MHD disk wind  as recently modeled by Panoglou et al. (2010, A\&A, submitted). 

\begin{acknowledgements}      
This work is based on archival data obtained with the Spitzer Space Telescope, which is operated by the Jet Propulsion Laboratory, California Institute of Technology under a contract with NASA. 
It was supported in part by the European Community's Marie Curie Actions - Human Resource and Mobility within the JETSET (Jet Simulations, Experiments and Theory) network under contract MRTN-CT-2004 05592. Financial contribution from contract ASI I/016/07/0 is also acknowledged.
\end{acknowledgements}

\bibliographystyle{aa} 
\bibliography{HH211} 

%



\clearpage

\end{document}